\documentclass[  amssymb, amsmath, aps,twocolumn,superscriptaddress, pra]{revtex4-2}
\usepackage{comment,amsmath,braket,bbold,dcolumn,csquotes,xspace,tikz,multirow,bbold}
\usepackage{graphicx}
\usepackage[nolist]{acronym}
\usepackage[normalem]{ulem}
\usepackage[utf8]{inputenc}

\usepackage[colorlinks=true,linkcolor=blue]{hyperref}%
\expandafter\ifx\csname package@font\endcsname\relax\else
 \expandafter\expandafter
 \expandafter\usepackage
 \expandafter\expandafter
 \expandafter{\csname package@font\endcsname}%
\fi
\usetikzlibrary{positioning}

\newcounter{note}
\renewcommand{\thenote}{\Roman{note}}

\makeatletter
\newcommand{\notefn}[1]{%
  \refstepcounter{note}%
  \hyperlink{note:\thenote}{\textsuperscript{\thenote}}%
  \insert\footins{%
    \reset@font\footnotesize
    \interlinepenalty\interfootnotelinepenalty
    \hypertarget{note:\thenote}{}%
    \hbox to 0.4em{{\hss\color{blue}\thenote}}%
    \vspace*{-\baselineskip}\setlength\belowdisplayshortskip{0pt}         
    \vspace{-0.15cm}
    #1\par
  }%
}
\makeatother

\newcommand{\inner}[2]{\braket{#1|#2}_{\text{KG}} }
\newcommand{\schro}{\text{Schr}\ddot{\text{o}}\text{dinger}}

\newcommand{\deriv}[3]{\frac{\partial^{#3} #1}{\partial #2^{#3} }}
\newcommand{\derivb}[2]{\partial^{#2}_{#1}}

\newcommand{\lad}[0]{\phi}

\newcommand{\psilad}[0]{\chi}

\newcommand{\bfr}[1]{{\bf #1}}

\begin{document}

\title{Algebra of quantum mechanics \textit{via} classical phonons. I: The $\schro$ equation as the 
Newtonian equation of motion and quantum observables as classical averages}

\author{Emmanuel Giner}%
\email{emmanuel.giner@lct.jussieu.fr}
\affiliation{Laboratoire de Chimie Th\'eorique, Sorbonne Universit\'e and CNRS, F-75005 Paris, France}

\begin{abstract}
The $\schro$ equation for a single spinless particle is formally obtained \textit{via} a classical phonon model, 
namely the Frenkel-Kontorova model.  
Starting from a one-dimensional lattice of coupled harmonic oscillators, we show that the continuous limit 
of the corresponding Newtonian equation of motion yields the Klein-Gordon equation for a real-valued field. 
By introducing a complex-valued change of variables mixing the real-valued displacement and velocity fields, and by separating fast and slow time scales, 
the Klein-Gordon equation is written as the $\schro$ equation within the non-relativistic limit. 
This complex change of variable also allows to rewrite classical global observables of the phonon field, such as the total energy or momentum, 
as the corresponding quantum observables. 
Additionally, we show that when a friction force is incorporated into the classical model, 
the corresponding Klein-Gordon equation can be rewritten as a $\schro$ equation with a non-Hermitian Hamiltonian. 
While the global approach is limited here to the non-relativistic regime and does not address the measurement problem, 
quantization or relativistic effects, 
it nonetheless illustrates how quantum algebra and complex-valued wave functions can be exactly reproduced using classical dynamics. 
The relativistic regime for a spinless particle and the link between commutators and Poisson brackets 
is addressed in the second part\cite{Giner-heisenberg} of this series. 
\end{abstract}

\maketitle

\section{Introduction}

Even without the probabilistic interpretation of the measurement problem, 
the algebra of quantum mechanics\cite{Dirac-book-30,VonNeumann-book-32} is radically different from that of classical mechanics. 
While the latter deals with the real-valued three-dimensional position and momentum vectors ruled by Newton's law of motion, the former  
uses the complex-valued $\schro$ scalar wave function, whose time evolution is governed by the $\schro$ equation. 
Then, observables are computed as expectation values of hermitian operators, which result in quadratic functionals of the $\schro$ wave function.  
These differences with classical mechanics have generated both philosophical and practical discussions, 
which are still undergoing\cite{Gibney-NAT-2025}. 
For instance, the status of whether the use of complex numbers in quantum mechanics is 
"intrinsic" or "only a handy mathematical tool" has been the subject of debates since the early foundations 
of the field (\textit{e.g.} by $\schro$ himself\cite{Schrodinger-26,Karam-AJP-20}) 
and found recently a renew of interests both 
theoretically\cite{McKagMosGis-PRL-09,HofWoo-arxiv-25,HitTruKamEppBru-PRL-26} 
and experimentally\cite{RenTriWeiLeTavGisAci-nat-21,Wu-PRL-22,Li-PRL-22}. 
More generally, there exists many derivations of the $\schro$ equation and the algebra 
of quantum mechanics which have been proposed in the literature, among which 
the Bohmian interpretation is certainly one of the more famous\cite{Bohm-PR-52}.  
Nevertheless, many other propositions have emerged 
using diffusion-based equations\cite{Moyal-MPCPS-49,Nelson-PR-66,Davies-JPa-89,OlaLapFig-AP-12,BeyPau-U-21,Yang-JMP-21}, 
information theory\cite{Reginatto-PRa-98},
Hamilton-Jacobi formalisms\cite{Grossing-FPL-04,Field-arxiv-12}, 
Feynman's path integral postulate\cite{Feynman-RVP-48,Derbes-aJP-96}, stochastic electrodynamics\cite{Derbes-aJP-96},
non-linear wave-like equations\cite{Kinsler2013Sep}, classical coupling through a bath\cite{BriRos-FP-01}, 
quaternionic framework\cite{DanLuc-E-20}, 
and more recently a generalization of non-Markovian stochastic processes\cite{Barandes-PP-25}. 
With respect to these works, we do not propose here an alternative formulation of quantum mechanics, nor a classical explanation of quantum phenomena. 
Instead, we simply show that several formal structures traditionally associated with non-relativistic quantum mechanics 
can be reproduced by the description of a classical system made of an infinite number of classical oscillators.

Another derivation of the algebra of non relativistic quantum mechanics 
can be obtained using the well established quantum field theory (QFT)
(see for instance lectures notes on QFT \cite{Greiner-book,QFTong,ZeeQFT} together 
with recent research works\cite{Padmanabhan-EPJC-18,BarGom-EPJC-21}). 
The starting point of these approaches are the classical Klein-Gordon fields (either real- or complex-valued) together 
with their associated Hamiltonian, which are then quantized by either promoting to operators the conjugated scalar fields 
and postulating the usual canonical non commutation relation between them, or by postulating the path integral 
formalism \textit{a la} Feynman based on the action associated to the field. 
The notion of particles (and anti particles in the case of complex fields) naturally appears as discrete 
excitations of these quantized fields operators, and one can then recover most of the usual algebra of many-particles 
non relativistic quantum mechanics for bosonic systems, such as the position and momentum operators 
whose commutator equals $i\hbar$. 
Therefore, although rigorous in its derivation from QFT, this approach nevertheless inherently uses the pillars of the quantum axioms, 
such as the canonical quantization, the Heisenberg representation or the Feynman path integral formalism. 

The aim of the present series of papers is to show how one can obtain the algebra of quantum mechanics  
for a single spinless particle, starting from the description of a classical system.  
The present paper focusses on the non relativistic regime using the Newtonian equation of motion, 
while the second paper\cite{Giner-heisenberg} targets the relativistic description through the Hamiltonian formalism. 
We therefore emphasize that we do not try here to solve the measurement problem (\textit{i.e} the non unitary evolution and the observed randomness of quantum mechanics), 
as the latter does not correspond straightforwardly to the quantum mechanical algebra such as the deterministic $\schro$ equation, 
and our aim is not neither to reinterpret quantum mechanics in the ontological sense. 
Also, while the model reproduces key features of quantum mechanics, 
it does not address spin, quantization schemes, or many-body systems, which remain the domain of QFT. 

Summarized in a few words, the approach developed in the present paper consists in starting from the Newtonian law of motion applied to a set of masses coupled by springs, 
from which the $\schro$ equation for a single spinless particle together with the algebra of non relativistic quantum mechanics 
are reproduced through some change of variables. 
We will also show in the second part\cite{Giner-heisenberg} of this series how the algebra of non commuting operators together with the Heisenberg picture 
can be obtained from the Poisson brackets of the Hamiltonian framework associated to the same classical model. 
Also, as the meaning of the word \textit{classical} can differ according to the context, we would like to clarify the meaning used here to avoid 
any misinterpretations of the aim of the present work.  

By \textit{classical} we mean here the usual laws of motion of \textit{classical mechanics}: 
trajectories of massive point particles in configuration space governed by second-order equations of motion,  
and which are submitted to external and/or internal forces or potentials. 
This is rather different from \textit{classical field theory} as it is usually defined 
in the context of QFT, \textit{i.e.} 
the system is described by a \textit{function} (or a vector) 
of space-time coordinates and not an \textit{operator} parametrized by space-time coordinates,  
and it obeys an equation of motion which can be retrieved from the stationary condition of an action functional. 
Considering the broad definition of an action functional, a \textit{classical field theory} 
is not actually related to Newtonian mechanics, thus highliting the ambiguity of the term \textit{classical}. 
For instance, the $\schro$, Dirac and Proca equations can be obtained from an action stationary principle 
and therefore fulfill the definition of a classical field theory, 
but they nevertheless do not explicitly rely on Newtonian mechanics. 
As we will here consider the limit of an infinite number of classical particles, 
we will obtain equations describing a continuous field,  
thus effectively obtaining a \textit{classical field theory},  
here real-valued Klein-Gordon fields. 
Nevertheless, despite this formal identification with a classical field theory, 
as the model used here starts with massive particles evolving according to Newtonian mechanics, 
the path to obtain the Schrodinger equation will be guided and motivated here 
only by \textit{classical mechanical} (\textit{i.e.} Newtonian mechanics) considerations. 
Therefore, the Schrödinger equation as obtained in this work should not be interpreted as a quantization of a classical field, 
but rather as a specific formulation of Newtonian dynamics for an infinite set of particles which effectively reproduces 
the bulk of the results of non relativistic quantum mechanics for a single particle. 
We would like to point out that we will be using a similar classical model than that proposed by Zee in the introduction 
of his famous book on QFT\cite{QFTZee}. Nevertheless, with respect to the latter, we will not quantize the classical model 
through the path integral formalism, but rather remain within a classical field framework, and push further 
the Newtonian description. It is also noteworthy that a similar model was recently used to perform a mechanical analogy 
between the Klein-Gordon equation and fluid mechanics\cite{HeiMaaMakPom-PF-22}. 
Nevertheless, with respect to the latter work, we avoid here to promote the real-valued Klein-Gordon field 
to a complex field in order to recover the $\schro$ equation. 
This perticuliar aspect is central to maintain the equivalence of the quantum description with 
that of the classical system, as the latter is intrinsically real-valued.

The main steps and organization of the present work can be summarized as follows. 
We start our derivations in Sec. \ref{sec:frenkont} 
by using a classical phonon-like system (the linearised Frenkel-Kontorova model\cite{FreKon-ZETF-38}) 
whose Newtonian equation of motion leads, in the continuous limit and with an appropriate choice of parameters, 
to the Klein-Gordon equation. 
The model used here is a generalization of the phonon model used by Zee\cite{QFTZee}, 
and we further analyze some key quantities related to the Klein-Gordon fields under the light of the classical nature of the model. 
From there, by performing a non relativistic limit in Sec. \ref{sec:schro_smooth}, we then obtain the $\schro$ equation, 
which is the main result of this work.   
Nevertheless, we do not follow the usual path connecting the $\schro$ and Klein-Gordon equations, 
which can be found in many textbooks on QFT (see for instance Refs. \onlinecite{Greiner-book,QFTong,QFTZee}).  
The main reason for that is that the usual path connecting the Klein-Gordon equation to the $\schro$ equation
necessarily involves a complex-valued Klein-Gordon field which is incompatible with our classical treatment, 
intrinsically real-valued. 
Instead, we perform a specific complex-valued change of variable 
(related to the two-component formalism of Feshback and Villars\cite{FesVil-RMP-58} and to the works of Mostafazadeh's on the Klein-Gordon equation\cite{Mostafazadeh-CQG-02}) 
allowing to rewrite the Klein-Gordon equation as a first-order differential equation similar to the $\schro$ equation. 
\notefn{The present author does not see yet a straightforward connection between 
the complex-valued change of variables used here and  
the concept of complex-valued "classical wave functions" $\Theta_n$ in phase space introduced by Sch{\ifmmode\ddot{o}\else\"{o}\fi}nberg\cite{Schonberg-NC-52,Schonberg-NC-53}. 
We nevertheless refer the reader to the very pedagogical and insightful historical perspective on the various works on Hilbert space formulation of classical mechanics by Barandes\cite{Barandes-EPJH-26}. } 
Then, we use a change of variable describing a slowly varying envelope which, under the non relativistic limit, 
transforms into the $\schro$ equation. 
Therefore, the complexness of the $\schro$ wave function and equation appear here as the result of 
a complex change of variable which is particularly well suited to separate two different time scales of oscillations. 
Based on this change of variable, we propose a classical picture of the wave function as an envelope of fast classical oscillations, 
without ontological implications (see Sec. \ref{seq:schro_kg_good}). 
Then, in Sec. \ref{sec:observables}, we show how the classical total quantities associated to the phonon system 
can be rewritten as the usual quantum expectation values. 
Eventually, we show in Sec. \ref{sec:nh_qm} that the addition of a classical friction force in the phonon model 
can be rewritten as a $\schro$ equation governed by a non hermitian Hamiltonian, similar to what is done to treat open systems 
in non relativistic quantum mechanics. 
Eventually, as the audience intended for this paper is mostly the non-relativistic community, 
we provide in the Appendix \ref{seq:schro_from_kg_usual} a brief summary of some results 
associated to the Klein-Gordon equation 
which are useful to connect the present work with this relativistic classical field theory. 
This includes the definitions of global quantities (such as the Lagrangian or momentum) 
together with why we shall not follow the usual path connecting the Klein-Gordon and $\schro$ equations. 

We would like to highlight that this paper is written mostly for physicists, 
and therefore one might not expect the full mathematical treatment when some limits are taken. 
For instance, the non relativistic limit $c\rightarrow \infty$ of the Klein-Gordon equation are here taken \textit{a la physicist}, 
but one can turn to more rigorous derivations of the latter in the mathematician literature 
(see for instance a Ref. \onlinecite{HasJiaSusVas-arxiv-25} and references therein). 
Lastly, we summarize in Table \ref{table} the different variables used here and their main features.

\begin{table*}[t]
\caption{\label{table}
        Summary of the different variables used in this work and their main features. 
        We also recall that $a$ is the lattice parameter in the discrete model, $\lambda_c=\hbar/mc$, and $\omega_0=mc^2/\hbar$. 
        Here, "fast" means that the typical time variation is on the order of $(\omega_0)^{-1}$, 
        while "slow" implies the typical time variations appears at a much larger time scale. 
        }
\begin{tabular}{l|c|l|c|c|c}
 Variables       & definition & \,\,\, unit & discrete/continuous &real/complex & fast/slow  \\
\hline
$u_{n}(t)$            & transverse displacement   & $[L]$  & discrete  & real  & fast \\
$\psi_{n}(t)$         & $\psi_{n}(t)=\sqrt{a} \lambda_c u_{n}(t)$ &  $[L]^{-1/2}$  & discrete  & real  & fast \\
$\psi(x,t)$         & $\lim_{a\rightarrow 0}\psi_{n}(t)$ &  $[L]^{-1/2}$  & continuous & real  & fast \\
$\psilad(x,t)$         & $\frac{1}{\sqrt{2}}\big(\psi(x,t) + \frac{i}{\omega_0}\dot{\psi}(x,t)\big)$ &  $[L]^{-1/2}$  & continuous & complex & fast \\
$\lad(x,t)$         & $e^{+i\omega_0 t}\psilad(x,t)$ &  $[L]^{-1/2}$  & continuous & complex & slow \\
\end{tabular}
\end{table*}

\section{The Klein-Gordon equation as a classical phonon model}\label{sec:frenkont}
The present section is dedicated to the study of the linearised Frenkel-Kontorova model\cite{FreKon-ZETF-38},  
qualitatively depicted in Fig. \ref{fig:FK_draw}, which allows to retrieve the Klein-Gordon equation in the continuous limit. 
This consists essentially in a generalization of the model used by Zee\cite{QFTZee} as it includes an external potential. 
Also, as we wish to perform a non relativistic limit, 
we chose here to write the model without using natural units as we want to easily keep track of $c$. 

\subsection{The one-dimensional Frenkel-Kontorova model}
\begin{figure}[t]
        \centering
        \includegraphics[width=\columnwidth]{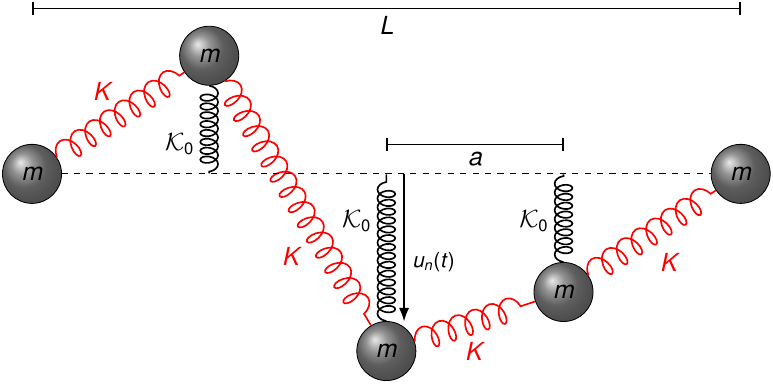}
        \caption{Pictorial representation of the Frenkel-Kontorova model used here. 
        The chain of length $L$ is made of $N$ particles  of mass $m$, 
        each of them are separated on the $x$ axis by the lattice parameter $a$. 
        The transverse displacement of the $n$-th mass is represented by the real number $u_n(t)$,  
        $\mathcal{K}_0$ is the strength of the "on-site" harmonic force and $K$ that of the "between-site" harmonic force.  }
        \label{fig:FK_draw}
\end{figure}

Consider a one-dimensional regular lattice of periodicity $a$ and length $L$. 
At each vertex ${\mathbf{X}_n=na \mathbf{e_x}, \,\, 1\le n \le N=L/a}$, lies a mass $m$, attached to a spring, 
characterized by a constant $\mathcal{K}_0$. The springs allow only for a transverse movement 
labelled $u_n(t)\in\mathbb{R}$, and ${\mathbf{U}(t)=\{ u_n(t),1\le n\le N \}}$ is the vector of all displacements at a given time $t$. 
We impose that the displacements vanish at the boundaries of the chain, \textit{i.e.} $u_1(t)=u_N(t)=0$. 
Also, each mass is connected to its two nearest neighbours, each one by a spring characterized by the constant $K$ 
and an equilibrium length of $a$. 
Therefore, the system is characterized by two types of forces: an "on-site" force, characterized by $\mathcal{K}_0$, 
and a "between-site" force, characterized by $K$ (see Fig. \ref{fig:FK_draw}). 

Within these notations, keeping only linear terms in the "between-site forces", 
the Newton's law of motion for the transverse displacement $u_n(t)$ is then 
\begin{equation}
 \label{eq:ds_0}
 \begin{aligned}
 & m \ddot{u}_n(t) - K \big(u_{n+1}(t) + u_{n-1}(t) - 2 u_n(t) \big) +  \mathcal{K}_0 u_n(t) =0.
 \end{aligned}
\end{equation}
In order to facilitate its continuous limit together with its identification with the usual wave function of 
non relativistic quantum mechanics, we introduce the rescaled displacement $\psi_n(t)$ as follows 
\begin{equation}\label{eq:rescaled}
 u_n(t) = \lambda_c \sqrt{a}{\psi}_n(t), \quad {\psi}(t) = \{ {\psi}_n(t),1\le n \le N\},
\end{equation}
where $\lambda_c=\hbar/mc$ is the Compton wave length, such that $\psi_n(t)$ has the unit of $L^{-1/2}$. 
Therefore, by linearity of Eq. \eqref{eq:ds_0}, ${\psi}_n(t)$ satisfies the same equation as $u_n(t)$.
Writing $K=\alpha/a^2$, Eq. \eqref{eq:ds_0} becomes 
\begin{equation}
 \label{eq:ds_1}
 \begin{aligned}
 & m \ddot{\psi}_n(t) - \alpha\frac{{\psi}_{n+1}(t) + {\psi}_{n-1}(t) - 2 {\psi}_n(t)}{a^2} +  \mathcal{K}_0 {\psi}_n(t) =0.
 \end{aligned}
\end{equation}
We can now perform the continuous limit of the system of coupled oscillators, 
which corresponds to $N\rightarrow \infty$, $a\rightarrow 0$, $Na =L$, 
such that  
\begin{equation}
 \begin{aligned}
  \lim_{\substack{a \rightarrow 0 \\ N\rightarrow \infty}} {\psi}_n(t) = {\psi}(x,t), \\
 \end{aligned}
\end{equation}
\begin{equation}
 \begin{aligned}
  \lim_{\substack{a \rightarrow 0 \\ N\rightarrow \infty}} \frac{{\psi}_{n+1}(t) + {\psi}_{n-1}(t) - 2 {\psi}_n(t)}{a^2} = \deriv{{\psi}(x,t)}{x}{2}.
 \end{aligned}
\end{equation}
Inserted in Eq. \eqref{eq:ds_1} and dividing by $m$ we obtain then 
\begin{equation}
 \begin{aligned}
  \ddot{\psi}(x,t) - \frac{\alpha}{m} \deriv{{\psi}(x,t)}{x}{2} +  \frac{\mathcal{K}_0}{m} {\psi}(x,t) =0.
 \end{aligned}
\end{equation}
For $\alpha= mc^2$ and $\mathcal{K}_0 = m^3c^4/\hbar^2$, we recover the Klein-Gordon equation 
in the one-dimensional case, namely 
\begin{equation}
 \label{eq:kg_one_d}
 \begin{aligned}
  \ddot{\psi}(x,t) - c^2 \deriv{{\psi}(x,t)}{x}{2} +  \omega_0^2  {\psi}(x,t) =0,
 \end{aligned}
\end{equation}
where we introduced the pulsation $\omega_0$ as
\begin{equation}\label{eq:omega_0}
 \omega_0 = \frac{mc^2}{\hbar}.
\end{equation}
One can therefore see the Klein-Gordon equation as the equation of motion for a classical vibrating string 
with an additional harmonic force trying to maintain the string at rest. 

One can then naturally extend the model by considering that 
the strength of the "on-site" spring is locally slightly modified with respect to $\omega_0$. 
More precisely, we consider now that the "on-site" frequency of the spring "$n$" is written as 
\begin{equation}\label{eq:delta}
 \omega_n^2 = \omega_0^2 \big( 1 + 2\frac{v_n}{mc^2}\big),
\end{equation}
where the factor $2$ has been put for further simplifications, and where $v_n$ has the unit of an energy.  
In the continuous limit, $v_n \rightarrow v(x)$, which can then be associated to the external potential $v(x)$, 
such that the Klein-Gordon equation becomes 
\begin{equation}
 \label{eq:kg_one_d}
 \begin{aligned}
  \ddot{\psi}(x,t) - c^2 \deriv{{\psi}(x,t)}{x}{2} +  \omega_0^2  (1+2\frac{v(x)}{mc^2}){\psi}(x,t) =0.
 \end{aligned}
\end{equation}
The case where $v(x)=0$ will naturally be referred to as the case of \textit{free} Klein-Gordon fields, 
although one has to keep in mind that, within this phonon model, there are harmonic forces acting on each point particle. 

We emphasize that the constants $m$, $c$ and $\hbar$ appearing throughout this paper arise from 
two choices: i) the spring constants are chosen so as to reproduce the Klein-Gordon equation and ii) 
$u_n$ is caled by the Compton length (see the definition of $\psi_n$ in Eq. \eqref{eq:rescaled}).

\subsection{Extension to higher spatial dimensions}
The extension of the model to the two-dimensional case is straightforward as it consists 
in a regular two-dimensional lattice of periodicity $a$, 
the masses being now indexed by two integers $(m,n)$, and $u_{m,n}(t)$ being the displacement orthogonal with respect to the lattice. 
Assuming that there are only nearest-neighbour couplings, and using the same choice of $\mathcal{K}_0$ and $K$ as previously, the equation of motion leads to  
\begin{equation}
 \label{eq:d2_0}
 \begin{aligned}
 &  \ddot{u}_{m,n}(t) +  \omega_0^2 u_{m,n}(t) \\
   &- c^2 \frac{u_{m,n+1}(t) + u_{m,n-1}(t) - 2 u_{m,n}(t) }{a^2} \\
   &- c^2 \frac{u_{m+1,n}(t) + u_{m-1,n}(t) - 2 u_{m,n}(t) }{a^2}  =0 .
 \end{aligned}
\end{equation}
With the rescaling $u_{m,n}(t)=\lambda_c a \psi_{m,n}(t)$ 
it yields in the continuous limit to $\psi_{m,n}(t)\rightarrow \psi(x,y,t)$, such that the equation becomes 
\begin{equation}
 \label{eq:kg_two_d}
 \begin{aligned}
  \ddot{\psi}(x,y,t) &+  \omega_0^2  {\psi}(x,y,t) \\ &- c^2 \big(\deriv{{\psi}(x,y,t)}{x}{2} + \deriv{{\psi}(x,y,t)}{y}{2}) =0.
 \end{aligned}
\end{equation}
This is the "mattress" model of Zee which was sketched by Zee in the introduction of his book on QFT\cite{ZeeQFT}.
Of course, adding a local fluctuation of the "on-site" strength of the springs as in Eq. \eqref{eq:delta} can be introduced and will lead in the continuous limit 
to the addition of an external potential $v(x,y)$ in Eq. \eqref{eq:kg_two_d}. 

Therefore, in the one- and two-dimensional cases, there is a one-to-one mapping between the phonon-like model and 
the Klein-Gordon equation describing real-valued fields. 
We can then straightforwardly interpret the Klein-Gordon field $\psi(x,y,t)$ as the local transverse displacement 
of this continuous phonon model, $\dot{\psi}(x,y,t)$ as the associated transverse velocity field, 
and the Klein-Gordon equation simply as the corresponding Newtonian equation of motion. 
The extension to the three-dimensional case is more conceptual as the transverse motion in three dimension 
would require a fourth spatial dimension. Nevertheless, one can simply imagine a medium whose local tension varies in a three-dimensional space, 
and whose equation of motion simply follows a Newtonian classical equation of motion. 
Also, the extension to the complex-valued Klein-Gordon fields can be trivially obtained: 
one can simply consider a complex-valued linear superposition of two classical real-valued fields, \textit{i.e.} 
$\Psi(x,t) = \psi_1(x,t) + i \psi_2(x,t)$, 
and thanks to the linearity of the equation of motion, such an abstract complex-valued field still satisfies the Klein-Gordon equation. 


As the extension of the formalism to the two- and three-dimensional case are straightforward, 
from thereon we focus on the one-dimensional case for the sake of simplicity of notations, 
except when dealing with the angular momentum which necessarily implies a three-dimensional space.

\subsection{Lagrangian and total energy}\label{eq:lag_htot}
The classical nature of the model allows to naturally retrieve the Lagrangian and total energy associated 
to the Klein-Gordon equation as usually given in classical field theory 
(see Appendix \ref{seq:schro_from_kg_usual} for a brief summary of such quantities). 
More precisely, using the discrete model, we can write the total kinetic energy as 
\begin{equation}
 T(\mathbf{U}(t)) = \frac{m}{2}\sum_{n=1}^N  \dot{u}_n(t)^2,
\end{equation}
or in terms of the vector of the scaled variables ${\mathbf{\Psi}(t) = \{ \psi_n(t) \}}$ as 
\begin{equation}\label{eq:kin_psin}
 T(\mathbf{\Psi}(t)) = \frac{mc^2}{2}\sum_{n=1}^N a \big(\frac{\dot{\psi}_n(t)}{\omega_0}\big)^2.
\end{equation}
Similarly, the total potential energy $V$ is written as 
\begin{equation}\label{eq:pot_psin}
 \begin{aligned}
 V(\mathbf{\Psi}(t))  =& \sum_{n=1}^Na \frac{1}{2}\big(mc^2+2v_n\big)\psi_n(t)^2 \\
            &+ \frac{\hbar^2}{2m}\sum_{n=1}^N a \frac{(\psi_n(t) - \psi_{n+1}(t))^2}{a^2}.
 \end{aligned}
\end{equation}
In the ${a\rightarrow 0}$ limit, we recover the continuum description, with ${\psi_n(t) \rightarrow \psi(x,t)}$, ${v_n \rightarrow v(x)}$ 
and ${\sum_{n}a \rightarrow \int dx}$. 
As a consequence, both the potential and kinetic energies of Eq. \eqref{eq:kin_psin} and Eq. \eqref{eq:pot_psin} 
become functionals of the continuous fields $\psi(x,t)$ and $\dot{\psi}(t)$, respectively, which explicitly reads 
\begin{equation}\label{eq:pot_psi}
 V[\psi(t)] = \frac{1}{2}\int dx \bigg(\big(mc^2+2v(x)\big)\psi(x,t)^2 + \frac{\hbar^2}{m}\big( \derivb{x}{}\psi(x,t)\big)^2\bigg),
\end{equation}
\begin{equation}\label{eq:kin_psin}
 T[\dot{\psi}(t)] = \frac{mc^2}{2}\int dx \bigg(\frac{\dot{\psi}(x,t)}{\omega_0}\bigg)^2.
\end{equation}
We can then define the Lagrangian density associated to the phonon model as the difference 
between the kinetic and potential energy densities, 
which explicitly reads 
\begin{equation}\label{eq:lagrange}
 \begin{aligned}
  \mathcal{L}  =& \frac{mc^2}{2}\bigg(\frac{\dot{\psi}(x,t)^2}{\omega_0^2} - \psi(x,t)^2\bigg) \\
          & - \frac{1}{2}\bigg(2v(x) \psi(x,t)^2 +  \frac{\hbar^2}{m} \big(\derivb{x}{}\psi(x,t)\big)^2\bigg).
 \end{aligned}
\end{equation}
Similarly, we can introduce the associated total energy by summing the total kinetic and potential energies
\begin{equation}\label{eq:cl_htot}
 \begin{aligned}
 H[\psi(t),\dot{\psi}(t)] & = T[\dot{\psi}(t)] + V[\psi(t)]  \\
            & \equiv mc^2\int dx \,\,\epsilon(\psi,\dot{\psi},x), 
 \end{aligned}
\end{equation}
where we introduced the energy density per unit of rest mass energy as follows 
\begin{equation}\label{eq:energy_rho}
 \begin{aligned}
 \epsilon(\psi,\dot{\psi},x) = 
 & \frac{1}{2} \bigg(\frac{\dot{\psi}(x,t)^2}{\omega_0^2} + \psi(x,t)^2\bigg) \\
 &+ \frac{1}{2mc^2}\bigg(2v(x) \psi(x,t)^2 +  \frac{\hbar^2}{m} \big(\derivb{x}{}\psi(x,t)\big)^2\bigg).
 \end{aligned}
\end{equation}
As the model uses only conservative forces, the total energy is conserved and therefore $H[\psi(t),\dot{\psi}(t)]$ is necessarily a constant of motion, \textit{i.e.}
\begin{equation}\label{eq:cst_energy}
 \frac{d}{dt}H[\psi(t),\dot{\psi}(t)] = 0.
\end{equation}
One can also notice that, in the case of $v(x)=0$, the Lagrangian and total energy of Eqs. \eqref{eq:lagrange} 
and \eqref{eq:cl_htot} are simply proportional to those found in the context of classical field theory,  
and we also recover without calculus that the integral of $T^{00}$ is a constant of motion 
(see Sec. \ref{seq:schro_from_kg_usual} for more details).

\subsection{The energy density as an inner product and a measure on Klein-Gordon fields}\label{sec:inner}
As the Klein-Gordon equation is linear, its solutions are stable by multiplication by an arbitrary scaling factor, 
which therefore introduces an ambiguity in the choice of these solutions. 
To remove such ambiguities one would need to define a norm and inner products on the spaces of Klein-Gordon solutions,  
and there exists various options to define inner products on the spaces of Klein-Gordon solutions 
(see for instance Ref. \onlinecite{MosZam-AP-06} and references therein). 
Here we choose an inner product which is not that of Mostafazadeh\cite{MosZam-AP-06}, 
but which will allow us to naturally define the localization of our phonon field as the energy barycentre (see Sec. \ref{seq:ehrenfest}). 
We briefly summarize in this section the various constraints associated to inner products on Klein-Gordon fields 
together with the explicit form and limit of the choice followed here. 

As the Klein-Gordon equation is second-order in time and linear, its solutions are uniquely determined by two initial conditions, 
$\psi(x,t=0)\equiv \psi_0(x)$ and  $\dot{\psi}(x,t=0)\equiv \dot{\psi}_0(x)$, which can be mapped into a two-dimensional vector 
$\Psi=(\psi_0, \dot{\psi}_0)$. 
Regarding now the inner products $\braket{\Psi_1|\Psi_2}$ on the space of the Klein-Gordon solutions, 
they must fulfill the following basic properties: i) as the vectors $\Psi$ are real-valued, $\braket{\Psi_1|\Psi_2}$ must be bilinear, 
ii) $\braket{\Psi_1|\Psi_2}$ must be symmetrical with respect to $\Psi_1$ and $\Psi_2$, 
and iii) $\braket{\Psi|\Psi}$ must be positive definite, \textit{i.e} $\braket{\Psi|\Psi} \ge 0$ and $\braket{\Psi|\Psi} = 0$  if and only if $\Psi=0$. 
Except for these generic properties of inner products, we require that $\braket{\Psi_1|\Psi_2}$ is time independent in order to 
be able to identify the space of solutions of the Klein-Gordon equation with the space of its initial conditions. 
Therefore, if $\Psi_1(t=0)$ and $\Psi_2(t=0)$ are two initial conditions of the Klein-Gordon equation, 
and $\Psi_1(t)$ and $\Psi_2(t)$ the corresponding solutions at time $t$, then the inner product must fulfill 
\begin{equation}
 \frac{d}{dt} \braket{\Psi_1(t)|\Psi_2(t)} =0. 
\end{equation}
Among the choices of inner product $\braket{\Psi_1|\Psi_2} $ fulfilling conditions i)-ii) and iii) together 
with time invariance, we choose here a definition which, when $\Psi_1=\Psi_2$, 
is proportional to the total energy as defined in Eq. \eqref{eq:cl_htot}.  
More precisely, we use the following inner product 
\begin{equation}\label{eq:inn_prod}
 \begin{aligned}
 &\inner{\Psi_1}{\Psi_2} \equiv \int dx 
  \frac{1}{2} \bigg(\frac{\dot{\psi}_1(x)}{\omega_0}\frac{\dot{\psi}_2(x)}{\omega_0} + \psi_1(x)\psi_2(x)\bigg) \\
 &+\! \frac{1}{2mc^2}\!\int\! dx \bigg(2v(x) \psi_1(x)\psi_2(x)\!  +\!   \frac{\hbar^2}{m} \derivb{x}{}\psi_1(x,t)\derivb{x}{}\psi_2(x,t)\!\bigg),
 \end{aligned}
\end{equation}
which fulfills all properties of a usual inner product, together with the time invariance.
It is also interesting to notice that the inner product can be written as follows 
\begin{equation}
 \inner{\Psi_1}{\Psi_2} = \frac{1}{2} \bigg(\frac{1}{\omega_0^2}\braket{\dot{\psi}_1|\dot{\psi}_2} + \braket{\psi_1|\psi_2}\bigg) 
 + \frac{1}{mc^2} \braket{\psi_1|\hat{h}\psi_2}, 
\end{equation}
where $\braket{f|g}$ is the usual $L^2$ inner product and $\hat{h}$ is the usual non relativistic Hamiltonian 
\begin{equation}\label{eq:egv_2}
 \hat{h} = -\frac{\hbar^2}{2m} \derivb{x}{2} + v(x).
\end{equation}
A sufficient condition on the potential $v(x)$ in order to guarantee the positivity of $\inner{\Psi}{\Psi}$ 
is that the ground state energy of $\hat{h}$ must be bounded from below by $-mc^2/2$. 
This condition can be illustrated in the case where $v(x)$ is the electron-nuclei Coulomb potential,  
which is an archetypal example of the non relativistic quantum calculations for molecules. 
As the ground state energy of an hydrogenoid atom of charge $Z$ is, in atomic units, of $-Z^2/2$, 
the positivity condition of $\inner{\Psi}{\Psi}$ fulfilled provided that $Z\le c \approx 137$. 
The latter condition suggests that the domain of validity of the inner product defined here 
is typically that of calculations where relativistic effects are negligible. 

As we will be frequently using the norm associated to a vector $\Psi=(\psi,\dot{\psi})$, we introduce the following notation 
\begin{equation}\label{eq:norm_psi}
 |(\psi,\dot{\psi})|^2 \equiv \inner{\Psi}{\Psi},
\end{equation}
such that the total energy of Eq. \eqref{eq:cl_htot} associated to a given solution of the Klein-Gordon equation $\Psi$ is then 
\begin{equation}
 H[\psi,\dot{\psi}] = mc^2 |(\psi,\dot{\psi})|^2 .
\end{equation}

Also, because of the definition of the energy density (\textit{i.e.} Eq. \eqref{eq:energy_rho}) we notice that 
\begin{equation}\label{eq:cond_pot}
 \text{if }v(x)>-\frac{mc^2}{2}\text{ then }
 \epsilon(\psi,\dot{\psi},x) >0 \,\,\forall\,\, (\psi,\dot{\psi}).
\end{equation}
The condition of Eq. \eqref{eq:cond_pot} can be illustrated in the case of an hydrogenoid Coulomb potential,  
where it is fulfilled when $|x|>Z\alpha\lambda_c$ where $\alpha\approx 7\times 10^{-3}$ is the fine structure constant. 
Therefore, for an electron of an hydrogenoid atom, 
the condition of Eq. \eqref{eq:cond_pot} is satisfied 
provided that the Coulomb potential is screened at the scale of $10^{-4}Z$, 
a regime in which the bare Coulomb potential is not physically relevant anyway owing to QED effects.

A consequence of the pointwise positivity of the energy density (\textit{i.e.} Eq. \eqref{eq:cond_pot}) 
is that for each $(\psi,\dot{\psi})$, the function $\epsilon(\psi,\dot{\psi},x)$ is a \textit{density} whose integral is time independent.
Therefore, similarly to a probability density, 
one can define the average of a function $f$ on the density associated to $(\psi,\dot{\psi})$, 
\textit{i.e.}
 \begin{equation}\label{eq:tilde_f}
 \begin{aligned}
\tilde{f}[\psi,\dot{\psi}]  & \equiv 
   \frac{\int dx\,\,f(x) \,\,\epsilon(\psi,\dot{\psi},x)}{\int dx \,\,\epsilon(\psi,\dot{\psi},x)}
 = \frac{\int dx\,\,f(x) \,\,\epsilon(\psi,\dot{\psi},x)}{|(\psi,\dot{\psi})|^2}. 
 \end{aligned}
 \end{equation}
The general form of Eq. \eqref{eq:tilde_f} will be used in Sec. \ref{seq:ehrenfest} to introduce the localization of the field. 

We end this section by emphasizing that, while the choice of the energy as the inner product is interesting 
to define the position of the field as the energy barycentre (see Sec. \ref{seq:ehrenfest}), 
it is nonetheless not Lorentz invariant, as opposed to the inner product introduced by Mostafazadeh\cite{Mostafazadeh-CQG-02}. 
Nevertheless, as the present paper focusses on the non relativistic limit, 
it is not a problem as the non invariance vanishes in the $c\rightarrow \infty$ limit. 
A more detailed discussion on other choices of inner products and densities within the relativistic regime 
is provided in the second part of this series of papers\cite{Giner-heisenberg}. 

\subsection{Global properties and Ehrenfest relations}\label{seq:ehrenfest}
Following the definition of averages over the density associated to the energy density (\textit{i.e.} Eq. \eqref{eq:tilde_f}), 
one can then estimate the localization of the field by introducing the energy barycentre as follows 
\begin{equation}\label{eq:av_pos}
 \begin{aligned}
 \tilde{X}[\psi,\dot{\psi}]  & = \frac{\int dx\,\,x \,\,\epsilon(\psi,\dot{\psi},x)}{\int dx \,\,\epsilon(\psi,\dot{\psi},x)} \\
                             & = \frac{\int dx\,\,x \,\,\epsilon(\psi,\dot{\psi},x)}{|(\psi,\dot{\psi})|^2}. 
 \end{aligned}
\end{equation}
Its time derivative can be computed after some algebra which yields 
\begin{equation}
 \frac{d}{dt} \tilde{X}[\psi,\dot{\psi}] = - \frac{ \lambda_c^2}{|(\psi,\dot{\psi})|^2 }\int dx \dot{\psi}(x,t) \derivb{x}{}\psi(x,t),
\end{equation}
suggesting the introduction of the total momentum 
\begin{equation}\label{eq:mom_cl}
 \begin{aligned}
 \tilde{P}[\psi,\dot{\psi}] =  -\frac{m \lambda_c^2}{|(\psi,\dot{\psi})|^2 }\int dx \dot{\psi}(x,t) \derivb{x}{}\psi(x,t),
 \end{aligned}
\end{equation}
such that one obtains the following Ehrenfest-like relation 
\begin{equation}
 \frac{d}{dt} \tilde{X} = \frac{\tilde{P}}{m}.
\end{equation} 
One can then obtain the time derivative of the momentum $P$ which, after some algebra, yields 
\begin{equation}
 \frac{d}{dt} \tilde{P}[\psi,\dot{\psi}] = - \frac{1}{|(\psi,\dot{\psi})|^2 }\int dx \psi(x,t)^2 \derivb{x}{}v(x),
\end{equation}
which suggests to introduce the averaged forces coming from the external potential $v(x)$ as 
\begin{equation}\label{eq:def_force}
 \tilde{F}[\psi,\dot{\psi}] = -\frac{1}{|(\psi,\dot{\psi})|^2 }\int dx \psi(x,t)^2 \derivb{x}{}v(x),
\end{equation}
such that we can write the second Ehrenfest-like relation
\begin{equation}
 \label{eq:ehrenfest}
 \frac{d}{dt} \tilde{P}[\psi,\dot{\psi}] = \tilde{F}[\psi,\dot{\psi}] .
\end{equation}
One can notice from Eq. \eqref{eq:ehrenfest} that, if the external potential is constant, the momentum is conserved \textit{i.e.}
\begin{equation}
\derivb{x}{}v(x)=0 \Rightarrow \frac{d}{dt} \tilde{P}[\psi,\dot{\psi}] =0,
\end{equation}
as in the case of the free Klein-Gordon field. 
Therefore, as long as the external potential does not break the translational invariance of the system, the momentum is conserved. 

In the three-dimensional case, we can introduce density of momentum as follows 
\begin{equation}
 \bfr{p}(\bfr{r},t) = -m \lambda_c^2  \dot{\psi}(\bfr{r},t) \nabla\psi(\bfr{r},t),
\end{equation}
such that one can rewrite the total momentum vector, which is the three-dimensional equivalent of Eq. \eqref{eq:mom_cl}, as follows 
\begin{equation}
 \begin{aligned}
 \tilde{\bfr{P}} &= -\frac{m \lambda_c^2}{|(\psi,\dot{\psi})|^2 }\int d^3\bfr{r} \,\dot{\psi}(\bfr{r},t) \,\,\nabla\psi(\bfr{r},t) \\
         &= \frac{1}{|(\psi,\dot{\psi})|^2 }\int d^3\bfr{r} \,\,\bfr{p}(\bfr{r},t). 
 \end{aligned}
\end{equation}
Therefore, one can introduce the local angular momentum density as 
\begin{equation}
\bfr{l}(\bfr{r},t) = \bfr{r}\times\bfr{p}(\bfr{r},t),
\end{equation}
such that the total angular momentum is defined as 
\begin{equation}\label{eq:angl_mom}
 \tilde{\bfr{L}} = \frac{1}{|(\psi,\dot{\psi})|^2 }\int d^3\bfr{r} \,\,\bfr{l}(\bfr{r},t).
\end{equation}
One can then compute its time derivative and, after some algebra, it yields that if $v(\bfr{r})=v(|\bfr{r}|)$, the angular momentum 
of Eq. \eqref{eq:angl_mom} is a constant of motion. 

We conclude this section by noticing that all the quantities derived here from the classical nature of the model 
are usually obtained from relativistic symmetry considerations in the context of classical field theory 
(see Appendix \ref{seq:schro_from_kg_usual} for a brief summary).

\subsection{Classical picture of the free Klein-Gordon equation and non relativistic limit}
\label{sec:interpret}
Based on the previous derivations, we now propose a qualitative analysis of the Klein-Gordon equation under the light of the 
classical nature of the model, and in terms of the two relevant parameters in our system which are $m$ and $c$. 
We will focus here on the free Klein-Gordon equation (\textit{i.e.} $\derivb{x}{}v(x)=0$) 
as the introduction of the external potential does not change the qualitative picture. 

The two key ingredients of the classical model are the "on-site" force proportional to $\omega_0^2 \psi(x,t)$, 
and the "between-site" force proportional to $c^2\derivb{x}{2} \psi(x,t)$. 
To qualitatively understand when one dominates on the other, we can compute the ratio of the "between-site" and the "on-site" 
potential energy densities at a given point $x$, which yields
\begin{equation}
 r \equiv 
 \lambda_c^2 \bigg(\frac{ \derivb{x}{}\psi(x,t)}{\psi(x,t)} \bigg)^2. 
\end{equation}
Therefore, asking that the "on-site" force dominates implies that $r \ll 1$, which is equivalent to 
\begin{equation}\label{eq:ratio}
 r \ll 1 \Leftrightarrow \bigg|\frac{ \derivb{x}{}\psi(x,t)}{\psi(x,t)}\bigg| \ll \lambda_c^{-1},
\end{equation}
which suggests that the term involving a spatial derivative is negligible in the Klein Gordon equation, 
which then becomes 
\begin{equation}
 \label{eq:kg_harm}
 \begin{aligned}
  \ddot{\psi}(x,t) +  \omega_0^2  {\psi}(x,t) =0.
 \end{aligned}
\end{equation}
The system described by Eq. \eqref{eq:kg_harm} is then made of a collection of uncoupled oscillator of pulsation $\omega_0$. 
As a consequence there is no wave propagation, as a wave packet will oscillate as an harmonic oscillator, but not propagate nor spread 
along the $x$ axis. 
On the other hand, in the $r \rightarrow \infty$, the "between-site force" dominates, and one obtains a 
usual wave equation where both the group and phase velocities are equal to $c$, \textit{i.e.}  
\begin{equation}
 \frac{1}{c^2}\derivb{t}{2}\psi(x,t) - \derivb{x}{2}\psi(x,t) = 0.
\end{equation}

Having established qualitatively the two limiting cases of the Klein-Gordon equation, 
we can interpret them in terms of the $m$ and $c$ parameters. 
Focussing first on the mass parameter, 
in the ${m\rightarrow\infty}$, the "on-site" force dominates (\textit{i.e.} $r\ll1$), which means that the mass prevent wave propagation, 
while the ${m\rightarrow 0}$ limit is equivalent to $r\gg1$, where waves propagate at the speed of light, suggesting photons. 
We can then recover a relatively intuitive understanding 
of the role of the mass : it consists in a parameter controlling the inertia of the system. 

Turning now to the $c$ parameter, the non relativistic limit translates into $r \ll 1$. 
According to Eq. \eqref{eq:ratio}, it implies that the typical spatial variations of $\psi(x,t)$ 
must be much larger than the Compton length. Assuming an oscillatory form $\psi(x,t)\propto e^{ikx}$ and using the De Broglie wave/particle relation,  
the non relativistic limit is equivalent to the condition that the usual quantum momentum 
is much smaller than associated relativistic momentum, \textit{i.e.} $\hbar k\ll mc$. 



\section{The $\schro$ equation as the Newtonian formalism describing the slowly-varying amplitude}
\label{seq:schro_kg_good}

As shown in Sec. \ref{sec:frenkont}, we can interpret the real-valued Klein-Gordon field $\psi(x,t)$ as a classical 
transverse displacement field of a continuous phonon model, whose time evolution follows simply Newtonian mechanics. 
In the present section we show how, starting from the Klein-Gordon equation, 
we can then obtain the $\schro$ equation, which appears as a rewriting of Newton's law of motion for 
the slowly varying envelope of the oscillations of the phonon field. 
Nevertheless, we will not follow the usual path 
to obtain the $\schro$ equation from a non relativistic limit of the Klein-Gordon equation, 
as it involves the assumption that the field $\psi(x,t)$ is intrinsically complex, 
which is incompatible with the classical picture (see Appendix \ref{seq:schro_from_kg_usual}). 

Instead, we will follow a path inspired by previous works re expressing the Klein-Gordon equation as 
an effective $\schro$ equation through a two-component framework\cite{FesVil-RMP-58,Mostafazadeh-CQG-02}. 
More precisely, we will use a specific case of the complex-valued change of variable introduced by Mostafazadeh\cite{Mostafazadeh-CQG-02}  
and show that in the non relativistic limit it leads indeed to the $\schro$ equation. 
Nevertheless, with respect to the work of Mostafazadeh or Feshback and Villars, here 
we will use only real-valued Klein-Gordon fields and exploit the classical nature of the phonon model previously introduced 
to propose a purely classical picture of the $\schro$ equation. 

\subsection{The $\schro$ wave function as a slowly-varying amplitude}\label{sec:schro_smooth}
In the absence of the external potential $v(x)$,  
we can notice that there are two different time scales in the Klein-Gordon equation: 
the shortest time scale corresponds to the "on-site" oscillations at frequency ${\omega_0}$, 
while the largest corresponds to the "between-sites" oscillations. 
As mentioned in Sec. \ref{sec:interpret}, if we neglect the "between-sites" oscillations 
(\textit{i.e.} the term proportional to $\derivb{x}{2}$), the system is simply a collection 
of uncoupled harmonic oscillators, each one vibrating with the same pulsation $\omega_0$. 
We will therefore use a change of variable allowing to compact the equation of motion of an harmonic oscillator, 
and apply it to the Klein-Gordon equation.  

Let us first consider the equation of motion of a simple classical harmonic oscillator 
\begin{equation}\label{eq:harm}
 \ddot{y}(t) + \omega_0^2 y(t) = 0.
\end{equation}
Although this is a second-order differential equation involving only real-valued quantities, 
it can be rewritten as the following complex-valued first-order differential equation 
\begin{equation}\label{eq:harm_comp}
 -i\dot{\psilad}(t) +\omega_0 \psilad(t) = 0,
\end{equation}
thanks to the following complex-valued change of variable 
\begin{equation}\label{eq:harm_psilad}
 \psilad(t) = \frac{1}{\sqrt{2}}\big( y(t) + \frac{i}{\omega_0}\dot{y}(t)\big).
\end{equation}
The general solutions of Eq. \eqref{eq:harm_comp} are then simply given by 
\begin{equation}\label{eq:psilad_harm}
 \psilad(t) = \psilad_0e^{-i\omega_0 t}, 
\quad \psilad_0 = \frac{1}{\sqrt{2}}\big( y(t=0) + \frac{i}{\omega_0}\dot{y}(t=0)\big) ,
\end{equation}
where the initial condition $\psilad_0$ gathers in a single complex number   
the two initial real-valued conditions on $y$ and $\dot{y}$. 
Another interesting property of the change of variable of Eq. \eqref{eq:harm_psilad} 
is that the total energy of the harmonic oscillator is then simply proportional to the norm of $\psilad$, \textit{i.e.}  
\begin{equation}
 \frac{1}{2}m\big( \dot{y}(t)^2 + \omega_0^2 y(t)^2) = m\omega_0^2 |\psilad(t)|^2.
\end{equation}
We can then apply this strategy to rewrite the Klein-Gordon equation with the following change of variable 
\begin{equation}\label{eq:psilad}
 \psilad(x,t) = \frac{1}{\sqrt{2}}\big( \psi(x,t) + \frac{i}{\omega_0}\dot{\psi}(x,t)\big),
\end{equation}
which corresponds to a special case of the general class of change of variables introduced 
by Mostafazadeh\cite{Mostafazadeh-CQG-02} (\textit{i.e.} with $\lambda=\omega_0^{-1}$ 
in the notations of Mostafazadeh's paper).  
Then, similarly to what is done in the case of a single classical harmonic oscillator, 
the Klein-Gordon equation is written in terms of $\psilad(x,t)$ and its complex conjugate $\psilad^*(x,t)$. 
Carrying the calculation, the Klein-Gordon equation is then written as 
\begin{equation}
 \label{eq:kg_psilad}
 \begin{aligned}
 -i\hbar \dot{\psilad}(x,t) -&\frac{\hbar^2}{2m}\derivb{x}{2}\psilad(x,t) 
 + v(x)\psilad(x,t) \\ 
& = -  \hbar\omega_0 \psilad(x,t) + \frac{\hbar^2}{2m}\derivb{x}{2}\psilad^*(x,t) - v(x)\psilad^*(x,t),  
 \end{aligned}
\end{equation}
which then resembles much more the $\schro$ equation. 
Nevertheless, with respect to the latter, Eq. \eqref{eq:kg_psilad} contains two additional terms : i) the term proportional to $\omega_0 \psilad(x,t)$ 
and ii) the terms involving $\psilad^*(x,t)$. 
As the term i) generates oscillations at the very large frequency $\omega_0$, we can assume that it corresponds to the shortest time scale. 
Therefore, we can absorb these rapid oscillations by performing a change of variable similar 
to the one employed usually to recover the $\schro$ equation from the Klein-Gordon equation (see Eq. \eqref{eq:psi_kg_usual}), \textit{i.e.}  
\begin{equation}\label{eq:def_lad}
 \psilad(x,t) = e^{-i\omega_0 t} \lad(x,t),
\end{equation}
where the function $\lad(x,t)$ describes typically the envelope of the oscillations. 
We emphasize here that, contrary to what is usually done in QFT textbooks to connect 
the Klein-Gordon and $\schro$ equations, the change of variables of Eq. \eqref{eq:def_lad} does not alter the 
classical interpretation. Indeed, in our framework the complex nature of the field $\psilad(x,t)$ 
is built in by construction, as it is defined as a combination of the real-valued transverse position and velocity fields 
(\textit{i.e.} Eq. \eqref{eq:psilad}). 
Inserting the form of $\psilad(x,t)$ written as in Eq. \eqref{eq:def_lad} in the writing of the Klein-Gordon equation as in Eq. \eqref{eq:kg_psilad},  
and multiplying by $e^{+i\omega_0 t}$, we obtain 
\begin{equation}\label{eq:kg_lad}
 \begin{aligned}
 -i\hbar \derivb{t}{} \lad(x,t) &-\frac{\hbar^2}{2m} \derivb{x}{2} \lad(x,t) + v(x)\lad(x,t) \\
&= e^{+2i\omega_0 t}\bigg(\frac{\hbar^2}{2m} \derivb{x}{2} \lad^*(x,t) + v(x) \lad^*(x,t)\bigg), 
 \end{aligned}
\end{equation}
which is simply a rewriting of the Klein-Gordon equation in terms of $\lad(x,t)$. 
We nevertheless notice that if right-hand side of Eq. \eqref{eq:kg_lad} vanishes, we obtain the $\schro$ equation. 
We now perform the non relativistic limit which corresponds to $\omega_0\rightarrow \infty$ in Eq. \eqref{eq:kg_lad}, 
and we will assume here that the time variation of $\lad(x,t)$ are negligible on the time scale of $\omega_0^{-1}$. 
More precisely, similarly to Eq. \eqref{eq:slow_1}, we assume now that 
\begin{equation}\label{eq:slow}
 \bigg|\frac{\derivb{t}{}\lad(x,t)}{\lad(x,t)}\bigg| \ll \omega_0,
\end{equation}
which corresponds to the well known slowly varying envelope approximation used for instance in electromagnetism. 
The form of Eq. \eqref{eq:slow} is equivalent to the assumption that the time average of $\lad(x,t)$ over a typical 
time scale of $\delta t= 2\pi\omega_0^{-1}$ is the same as  $\lad(x,t)$, \textit{i.e.}
\begin{equation}\label{eq:time_av}
 \frac{1}{\delta t} \int_{t}^{t+\delta t} ds \lad(x,s) \approx \lad(x,t).
\end{equation}
Therefore if we perform a time average of the Klein-Gordon equation as written in Eq. \eqref{eq:kg_lad},
we obtain, under the slowly varying envelope approximation of Eq. \eqref{eq:slow}, 
that the left-hand side is simply 
\begin{equation}
 \begin{aligned}
 \frac{1}{\delta t}\int_{t}^{t+\delta t} ds 
 &\big(-i\hbar \derivb{s}{} -\frac{\hbar^2}{2m} \derivb{x}{2} + v(x) \big) \lad(x,s) \\
\approx &\big(-i\hbar \derivb{t}{} -\frac{\hbar^2}{2m} \derivb{x}{2} + v(x) \big)  \lad(x,t),
 \end{aligned}
\end{equation}
while the right-hand side of Eq. \eqref{eq:kg_lad} yields 
\begin{equation}\label{eq:right-hand}
 \begin{aligned}
 &\frac{1}{\delta t} \int_{t}^{t+\delta t} ds e^{+2i\omega_0 s} 
\big(\frac{\hbar^2}{2m} \derivb{x}{2} + v(x)\big)\lad^*(x,s) \\
\approx &\frac{1}{\delta t} 
\big(\frac{\hbar^2}{2m} \derivb{x}{2} + v(x)\big)\lad^*(x,s) \int_{t}^{t+\delta t} ds e^{+2i\omega_0 s}= 0.
 \end{aligned}
\end{equation}
Therefore, within the slowly-varying envelope approximation, which corresponds to the non relativistic approximation, 
the right-hand side of Eq. \eqref{eq:kg_lad} vanishes and one recovers the $\schro$ equation. 

We conclude this section by connecting the slowly varying envelop approximation and the non relativistic limit. 
Consider the case of an eigenfunction of the $\schro$ equation for a free particle  
\begin{equation}
 \lad(x,t) \propto e^{ikx} e^{-it\frac{\hbar}{2m}k^2},
\end{equation}
then the slowly varying envelope approximation of Eq. \eqref{eq:slow} is simply equivalent to 
\begin{equation}
 \begin{aligned}
 &\hbar k \ll mc,\\
 \end{aligned}
\end{equation}
which is precisely the non relativistic approximation. 
Therefore, provided that the solutions of the $\schro$ equation do not have large momentum components 
compared to $mc$, the slowly varying approximation is valid.

\subsection{Eigenvectors and eigenvalues of the non relativistic Hamiltonian}
Another important aspect of non relativistic quantum mechanics are the eigenvectors and eigenvalues of the 
non relativistic Hamiltonian. We will show here that they can be thought as the normal modes of the classical phonon model. 

We begin by noticing that we can rewrite 
the Klein-Gordon equation of Eq. \eqref{eq:kg_one_d} in terms of the usual non relativistic Hamiltonian operator 
defined in Eq. \eqref{eq:egv_2}, which yields 
\begin{equation}
 \label{eq:egv_1}
 \begin{aligned}
 &  \ddot{\psi}(x,t) +\omega_0^2 (1 + \frac{2}{mc^2} \hat{h})\psi(x,t) = 0.
 \end{aligned}
\end{equation}
Considering now the eigenvectors of the non relativistic Hamiltonian
\begin{equation}\label{eq:egv_3}
 \hat{h} \phi_n(x) = \hbar \omega_n \phi_n(x), \quad \omega_n \in \mathbb{R},
\end{equation}
as the set of $\phi_n$ forms a complete basis, it can be used to decompose the Klein-Gordon field $\psi(x,t)$ as 
\begin{equation}\label{eq:egv_4}
 \psi(x,t) = \sum_{n} c_n(t) \phi_n(x).
\end{equation}
If one is interested with solutions $\psi(x,t)$ vanishing on the boundaries, the eigenvectors of $\hat{h}$ 
can be chosen real-valued, and therefore the $c_n(t)$ are also real-valued.  
Using the spectral decomposition of Eq. \eqref{eq:egv_4}, 
the Klein-Gordon equation as written in Eq. \eqref{eq:egv_1} becomes a set of uncoupled equations of motions for harmonic oscillators 
\begin{equation}\label{eq:egv_5}
 \ddot{c}_n(t) +\Omega_n^2 c_n(t) = 0,
\end{equation}
where we defined the pulsation associated to an eigenvector $n$ as follows 
\begin{equation}\label{eq:egv_6}
 \Omega_n^2 = \omega_0^2\big(1 + 2\frac{\omega_n}{\omega_0}\big) .
\end{equation}
Such a situation is typical from the normal modes of coupled oscillators,  
and therefore the eigenvectors of $\hat{h}$ appear as the normal modes of the system. 

We can also perform the same change of variables used in the case of the harmonic oscillator 
(see Eq. \eqref{eq:harm_psilad}) and introduce 
\begin{equation}\label{eq:egv_7}
 \tilde{c}_n(t) = \frac{1}{\sqrt{2}}\big( c_n(t) + \frac{i}{\Omega_n}\dot{c}_n(t)\big),
\end{equation}
such that the harmonic equation of motion of Eq. \eqref{eq:egv_5} is transformed as in Eq. \eqref{eq:harm_comp}, 
which yields 
\begin{equation}\label{eq:harm_comp_eigv}
 -i\dot{\psilad}_n(t) +\Omega_n \psilad_n(t) = 0.
\end{equation}
If one performs now the change of variable allowing to absorb the fast oscillations at pace $\omega_0$ 
\begin{equation}
 a_n(t) = e^{+i\omega_0 t} \tilde{c}_n(t), 
\end{equation}
and postulates a slowly varying envelope for ${a}_n(t)$ (\textit{i.e.} similar to Eq. \eqref{eq:slow}) 
together with an expansion at large $c$ of $\Omega_n$ 
\begin{equation}
 \Omega_n \approx \omega_0 + \omega_n + o(c^{-2}),
\end{equation}
inserted into Eq. \eqref{eq:harm_comp_eigv} and multiplied by $\hbar$, it yields o 
\begin{equation}
 i\hbar \dot{a}_n(t)  =\hbar \omega_n a_n(t) ,
\end{equation}
which is the expression of the time-dependent $\schro$ equation on the $\{\phi_n\}$ basis.

\section{Observables in non relativistic quantum mechanics}\label{sec:observables}
In the present section we show how the classical global observables defined in Sec. \ref{seq:ehrenfest}, 
such as the inner product, the energy barycentre or the total momentum, are written, 
in the non relativistic limit, as the usual expectation values 
of the corresponding hermitian operators over the $\schro$ wave function. 
We highlight briefly here the main procedure that will be used along these derivations:  
i) consider a classical quantity computed as an integral involving a quadratic form of $(\psi,\dot{\psi})$, 
ii) re express it in terms of the slow-motion envelope variables $(\lad,\lad^*)$, 
iii) perform a non relativistic limit together with the short-time average as before. 
Then, all terms which are not proportional to  $\lad \times \lad^*$ will 
vanish because they will necessarily involve $e^{\pm 2i\omega_0 t }$ which yield zero over the time average. 
The surviving terms expressed in terms of $(\lad,\lad^*)$ is the result of non relativistic quantum mechanics.

\subsection{The non relativistic norm and inner product}\label{sec:norm_energy}
We begin by decomposing the norm associated to the Klein-Gordon fields $(\psi,\dot{\psi})$ 
(see Eq. \eqref{eq:norm_psi}) as follows 
\begin{equation}\label{eq:decomposition_n}
 |(\psi,\dot{\psi})|^2 = \mathcal{N}[\psi,\dot{\psi}] + \frac{1}{mc^2}\mathcal{V}[\psi,\dot{\psi}],
\end{equation}
with 
\begin{equation}
 \begin{aligned}\label{eq:norm_1}
  \mathcal{N}[\psi,\dot{\psi}] & = \frac{1}{2}\int dx \bigg(\frac{\dot{\psi}(x,t)^2}{\omega_0^2} + \psi(x,t)^2\bigg), 
 \end{aligned}
\end{equation}
and 
\begin{equation}\label{eq:tot_e_lad_0}
 \begin{aligned}
  \mathcal{V}[\psi,\dot{\psi}]  & =  \int dx v(x) \psi(x,t)^2 + \frac{\hbar^2}{2m}\int dx \big(\derivb{x}{}\psi(x,t)\big)^2 \\
                 & = \int dx \psi(x,t) \hat{h} \psi(x,t).
 \end{aligned}
\end{equation}
In the non relativistic limit and provided that the spectrum of the non relativistic Hamiltonian is bounded by $-mc^2/2$, 
the term $\mathcal{V}/mc^2$ vanishes, such that 
\begin{equation}\label{eq:norm_nr}
 \lim_{c\rightarrow \infty} |(\psi,\dot{\psi})|^2 = \mathcal{N}[\psi,\dot{\psi}] . 
\end{equation}
We can then express the term  $\mathcal{N}[\psi,\dot{\psi}] $ in terms of 
the high-frequency complex-valued variables $(\psilad,\psilad^*)$, 
\begin{equation}
 \begin{aligned}\label{eq:norm_tmp}
  \mathcal{N}[\psilad,\psilad^*]  & = \int dx |\psilad(x,t)|^2 \\
 \end{aligned}
\end{equation}
and eventually in terms of the slow-motion complex-valued envelope variables $(\lad,\lad^*)$, which yields 
\begin{equation}
 \begin{aligned}\label{eq:norm_0}
  \mathcal{N}[\lad]  & = \int dx |\lad(x,t)|^2 \\
                     & = \braket{\phi|\phi},
 \end{aligned}
\end{equation}
where we recall that $\braket{f|f}$ is the usual $L^2$ inner product. 
Therefore, we obtain that 
\begin{equation}\label{eq:norm}
  \lim_{c\rightarrow \infty} |(\psi,\dot{\psi})|^2   = \braket{\lad|\lad}.
\end{equation}
Remembering that $\lad(x,t)$ becomes the $\schro$ wave function in the $c\rightarrow \infty$ limit, 
we therefore see that the norm associated to the Klein-Gordon fields becomes, in the non relativistic limit, 
the usual definition of the norm of the wave function in non relativistic quantum mechanics. 

The same conclusions can be generalized to the inner product defined in Sec. \ref{sec:inner}. 
We can rewrite the inner product of Eq. \eqref{eq:inn_prod} using the $(\lad,\lad^*)$ basis 
\begin{equation}\label{eq:inn_prod_lad}
 \begin{aligned}
 &\inner{\Psi_1}{\Psi_2} = \braket{\lad_1|\lad_2} \\
 &+ \frac{1}{2mc^2}\!\int\! dx \big( e^{-2i\omega_0 t} \phi_1(x) \hat{h} \phi_2(x)
 +  e^{+2i\omega_0 t} \phi_1^*(x) \hat{h} \phi_2^*(x) \big),
 \end{aligned}
\end{equation}
such that in the non relativistic limit we obtain the usual $L^2$ scalar product of non relativistic quantum mechanics 
\begin{equation}
 \lim_{c\rightarrow \infty} \inner{\Psi_1}{\Psi_2} =\braket{\lad_1|\lad_2}.
\end{equation}

One can also sketch several remarks motivated by the classical picture of the phonon model. 
The physical content of $\mathcal{N}$ is the kinetic energy and "on-site" dominant potential energy,  
while $\mathcal{V}$ gathers the rest of the potential energy, which is composed of the "between-site" 
and weak "on-site" potential energies. 
Therefore, $\mathcal{N}$ is typically associated to high-frequency oscillations, while $\mathcal{V}$ 
follows a slower time variation. 
For a finite value of $c$, both $\mathcal{N}$ and $\mathcal{V}$ fluctuate in time, 
but as the sum of the two terms is constant, it implies that there is an energy flow between both quantities. 
Nevertheless, in the non relativistic limit, both $\mathcal{N}$ and $\mathcal{V}$ are constants of motion of the $\schro$ equation, 
and therefore the energetic flux between these two terms vanishes. 
The latter situation is typical for a system where a time scale becomes negligible, 
which is the case here on time scale on the order of $\omega_0^{-1}$. 
Eventually, we highlight that the usual $L^2$ norm of the $\schro$ wave function 
can be thought, within the present framework, as the energy density associated to the high-frequency oscillations of the classical phonon system.

\subsection{The non relativistic energy}\label{sec:norm_energy}
Focussing now on the total energy of the system defined in Eq. \eqref{eq:cl_htot}, 
we can notice that such quantity is not invariant with respect to a scaling factor of both $(\psi,\dot{\psi})$. 
We therefore propose here to study the normalized energy as follows 
\begin{equation}\label{eq:tot_e_bis}
 \tilde{H}  = \frac{H[\psi,\dot{\psi}]}{|(\psi,\dot{\psi})|^2},
\end{equation}
which, because of the definition of $|(\psi,\dot{\psi})|^2$ (see Eq. \eqref{eq:norm_psi}), is nothing but $mc^2$. 
We can nevertheless write it explicitly as follows 
\begin{equation}\label{eq:tot_e_1}
 \tilde{H}  = mc^2 \frac{\mathcal{N} + \frac{1}{mc^2}\mathcal{V}}{\mathcal{N} + \frac{1}{mc^2}\mathcal{V}},
\end{equation}
and in the non relativistic limit, as the denominator of Eq. \eqref{eq:tot_e_1} tends towards $\mathcal{N}$, we obtain 
\begin{equation}\label{eq:tot_e_2}
 \begin{aligned}
 \lim_{c \rightarrow \infty }\tilde{H}  & = \lim_{c\rightarrow \infty}\frac{\mathcal{V}}{\mathcal{N}} + \lim_{c\rightarrow \infty}mc^2. 
 \end{aligned}
\end{equation}
Therefore, Eq. \eqref{eq:tot_e_2} is the typical expression of a non relativistic energy: 
the total energy is made of the internal energy to which the rest mass energy is simply added with no coupling 
between these two terms. 
Nevertheless, as the rest mass energy diverges in the $c\rightarrow \infty$ limit, one can then define the 
non relativistic energy as 
\begin{equation}\label{eq:e_nr_0}
 \begin{aligned}
 E^{\text{nr}}[\psi,\dot{\psi}] & = \lim_{c \rightarrow \infty}\tilde{H} - mc^2 \\
                                & = \lim_{c \rightarrow \infty}\frac{\mathcal{V}}{\mathcal{N}}.
 \end{aligned}
\end{equation}
In order to take the non relativistic limit appearing in Eq. \eqref{eq:e_nr_0}, we first express $\mathcal{V}$ in terms of the 
slow-motion complex-valued envelope variables $(\lad,\lad^*)$, which yields 
\begin{equation}\label{eq:tot_e_lad}
 \begin{aligned}
  \mathcal{V} 
                 & = \frac{1}{2}\int dx \big(\lad^*(x,t) \hat{h} \lad(x,t) + \lad(x,t) \hat{h} \lad^*(x,t) \\
  &+e^{-2i\omega_0t}\lad(x,t) \hat{h} \lad(x,t) + e^{+2i\omega_0t}\lad^*(x,t) \hat{h} \lad^*(x,t)  \big).\\
 \end{aligned}
\end{equation}
Then, in the non relativistic limit, the oscillating terms $e^{\pm 2i\omega_0 t}$ vanish 
and, by using the hermitian character of $\hat{h}$, the expression of $ \mathcal{V}$ as in Eq. \eqref{eq:tot_e_lad} then becomes simply 
\begin{equation}\label{eq:v_nr}
 \begin{aligned}
 \lim_{c\rightarrow \infty}\mathcal{V} & =\int dx \lad^*(x,t) \hat{h} \lad(x,t).
 \end{aligned}
\end{equation}
As the norm $\mathcal{N}$ is the $L^2$ norm of the $\phi$ (see Eq. \eqref{eq:norm_0}), we can then write the expression of Eq. \eqref{eq:e_nr_0} as follows
\begin{equation}\label{eq:e_nr}
 \begin{aligned}
 E^{\text{nr}}[\psi,\dot{\psi}] & = \frac{\int dx \lad^*(x,t) \hat{h} \lad(x,t)}{\int dx |\lad(x,t)|^2 } \\
                                & = \frac{\braket{\phi|\hat{h}|\phi}}{\braket{\phi|\phi}},
 \end{aligned}
\end{equation}
which is the expression of the kinetic and potential energy in the non relativistic quantum mechanics. 

We would like to conclude this section by commenting on the interpretation of the Laplacian operator, 
which is associated to the kinetic energy in the usual presentation of non relativistic quantum mechanics. 
In the present framework, the Laplacian operator comes from a purely potential energy term,   
as it translates the incompressibility of the phonon model. 
Therefore, one can have a classical picture of the quantum confinement energy as simply the fact that confining 
a particle leads necessarily to a tension in the phonon field, and therefore an increase of the potential energy. 
This contrasts with the usual result of non relativistic quantum mechanics which essentially states that the confining a particle necessarily 
induces an increase in \textit{kinetic} energy. 


\subsection{Position operator}\label{sec:position}
We proceed with the energy barycentre, which estimates the localisation of the field. 
Its expression as given in Eq. \eqref{eq:av_pos} can be rewritten as 
\begin{equation}
 \begin{aligned}
 \tilde{X}[\psi,\dot{\psi}] & = \frac{X[\psi,\dot{\psi}]}{|(\psi,\dot{\psi})|^2} + \frac{\delta X[\psi,\dot{\psi}]}{|(\psi,\dot{\psi})|^2}, \\
 \end{aligned}
\end{equation}
where 
\begin{equation}
 \begin{aligned}\label{eq:x_nr}
 X[\psi,\dot{\psi}] = \frac{1}{2}\int dx\,\,x\,\,\bigg(\frac{\dot{\psi}(x,t)^2}{\omega_0^2} + \psi(x,t)^2\bigg),
 \end{aligned}
\end{equation}
and 
\begin{equation}
 \begin{aligned}
 \delta X[\psi,\dot{\psi}] = \!\frac{1}{2mc^2} \!\int dx\,\,x \bigg(\!v(x) \psi(x,t)^2 
   +  \frac{\hbar^2}{m} \big(\derivb{x}{}\psi(x,t)\big)^2\!\bigg).
 \end{aligned}
\end{equation}
Nevertheless, in $c\rightarrow \infty$ limit, the term $\delta X[\psi,\dot{\psi}]$ vanishes as it is proportional 
to $c^{-2}$, such that the non relativistic position is obtained as 
\begin{equation}
 \begin{aligned}
 \label{eq:def_x_cl}
 \lim_{c\rightarrow \infty} \tilde{X}[\psi,\dot{\psi}] = \frac{X[\psi,\dot{\psi}]}{\braket{\lad|\lad}} ,
 \end{aligned}
\end{equation}
where we replaced $|(\psi,\dot{\psi})|^2$ by $\braket{\lad|\lad}$, which is its non relativistic limit (see Eq. \eqref{eq:norm}). 
We can then express the integrand of Eq. \eqref{eq:def_x_cl} in terms of the low-frequency envelope $(\lad,\lad^*)$ variables, 
which yields simply 
\begin{equation}
 X[\lad,\lad^*] = \int dx \,\,x\,\,|\lad(x,t)|^2,
\end{equation}
where we emphasize that $\lad$ becomes the $\schro$ wave function in the $c\rightarrow \infty$ limit. 
We therefore recover the usual non relativistic result of the position expectation value, \textit{i.e.} 
\begin{equation}
 \begin{aligned}
 \lim_{c\rightarrow \infty}\tilde{X}[\psi,\dot{\psi}] & = \frac{\braket{\phi|\hat{X}|\phi}}{\braket{\phi|\phi}},
 \end{aligned}
\end{equation}
with $\hat{X}=x$.

\subsection{Momentum operator}\label{sec:momentum}
The classical total momentum of the system as defined in Eq. \eqref{eq:mom_cl} can be expressed in terms 
of the couple $(\psilad,\psilad^*)$  as follows 
\begin{equation}\label{eq:tot_p_lad}
 \begin{aligned}
 \tilde{P} = \!\!-\frac{i\hbar}{2|(\psi,\dot{\psi})|^2}&\int\! dx \bigg(\!\psilad^*(x,t) \derivb{x}{}\psilad(x,t) 
  - \psilad(x,t) \derivb{x}{}\psilad^*(x,t) \\
 &- \psilad(x,t) \derivb{x}{}\psilad(x,t) +\psilad^*(x,t) \derivb{x}{}\psilad^*(x,t)\bigg).
 \end{aligned}
\end{equation}
We highlight here that the purely imaginary number $i$ appears in Eq. \eqref{eq:tot_p_lad} because 
the total classical momentum of Eq. \eqref{eq:mom_cl} depends linearly on the variable $\dot{\psi}$, 
which is proportional to the imaginary component of $\psilad$. 
Remembering that $\psilad$ vanishes on the boundaries, we can use 
\begin{equation}\label{eq:int_tric}
 \int_{a}^{b} dx f(x)\derivb{x}{}f(x) = 0 \quad \text{if }f(a)=f(b)=0, 
\end{equation}
such that only the terms involving the product of $\psilad^*$ and $\psilad$ in Eq. \eqref{eq:tot_p_lad} are left.  
As a consequence, Eq. \eqref{eq:tot_p_lad} is written as 
\begin{equation}\label{eq:tot_p_lad_2}
 \tilde{P}\! =\! -\frac{i\hbar}{2|(\psi,\dot{\psi})|^2}\int\! dx \bigg(\psilad^*(x,t) \derivb{x}{}\psilad(x,t) - \psilad(x,t) \derivb{x}{}\psilad^*(x,t) \!\bigg),
\end{equation}
which, by integrating by part, yields 
\begin{equation}\label{eq:tot_p_lad_3}
 \tilde{P} = -\frac{i\hbar}{|(\psi,\dot{\psi})|^2}\int dx \psilad^*(x,t) \derivb{x}{}\psilad(x,t).  
\end{equation}
Expressing the momentum of Eq. \eqref{eq:tot_p_lad_3} in terms of the slowly varying envelope $(\lad,\lad^*)$ simply leads to 
\begin{equation}\label{eq:tot_p_psilad_3}
 \tilde{P} = -\frac{i\hbar}{|(\psi,\dot{\psi})|^2}\int dx \lad^*(x,t) \derivb{x}{}\lad(x,t).  
\end{equation}
Therefore, in the non relativistic regime, $|(\psi,\dot{\psi})|^2$ becomes the usual $L^2$ norm of the wave function and Eq. \eqref{eq:tot_p_psilad_3} 
becomes then 
\begin{equation}\label{eq:tot_p_psilad_4}
 \lim_{c\rightarrow \infty} \tilde{P} = \frac{\braket{\phi|\hat{P}|\phi}}{\braket{\phi|\phi}},
\end{equation}
with $\hat{P}=-i\hbar \derivb{x}{}$, 
which is the usual definition of the momentum in non relativistic quantum mechanics. 

\subsection{The angular momentum}\label{sec:angular}
The classical total angular momentum functional defined as Eq. \eqref{eq:angl_mom} can be expressed in terms of the $(\psilad,\psilad^*)$ variables as 
\begin{equation}\label{eq:angl_mom_1}
 \begin{aligned}
 \tilde{\textbf{L}} = \frac{i\hbar}{2|(\psi,\dot{\psi})|^2}\int d^3\bfr{r} &\big(\psilad(\bfr{r},t)-\psilad^*(\bfr{r},t)\big) \\
 &\bfr{r} \times \nabla \big(\psilad(\bfr{r},t)+\psilad^*(\bfr{r},t)\big).
 \end{aligned}
\end{equation}
Once more, the purely imaginary number $i$ appears in Eq. \eqref{eq:angl_mom_1} because the angular momentum of 
Eq. \eqref{eq:angl_mom} depends linearly on the variable $\dot{\psi}$. 
The integrand of Eq. \eqref{eq:angl_mom_1} can be expanded as 
\begin{equation}\label{eq:angl_mom_bis}
 \begin{aligned}
  \int d^3\bfr{r} \,\,\,\bfr{r} \times \bigg(  &\psilad(\bfr{r},t) \nabla \psilad^*(\bfr{r},t) - \psilad^*(\bfr{r},t) \nabla \psilad(\bfr{r},t) \\
 +&  \psilad(\bfr{r},t) \nabla \psilad(\bfr{r},t) - \psilad^*(\bfr{r},t) \nabla \psilad^*(\bfr{r},t) \bigg).
 \end{aligned}
\end{equation}
By using the fact that $\psilad$ vanishes on the boundary, one can use Eq. \eqref{eq:int_tric} and notice that 
\begin{equation}
 \begin{aligned}
 &\int d^3\bfr{r} \,\, \psilad(\bfr{r},t) \,\,\bfr{r} \times \nabla \psilad(\bfr{r},t) = 0, \\
 \end{aligned}
\end{equation}
such that, by integration by part, the angular momentum of Eq. \eqref{eq:angl_mom_1} is written as 
\begin{equation}\label{eq:angl_mom_final_1}
 \begin{aligned}
 \tilde{\textbf{L}} = -\frac{i\hbar}{|(\psi,\dot{\psi})|^2}\int d^3\bfr{r} \,\,\psilad^*(\bfr{r},t) \,\, \bfr{r} \times \nabla \psilad(\bfr{r},t).
 \end{aligned}
\end{equation}
As before, we now perform the non relativistic limit and switch to the $(\lad,\lad^*)$ variable, and we obtain  
\begin{equation}\label{eq:angl_mom_final}
 \begin{aligned}
 \tilde{\textbf{L}} = 
 \frac{-i\hbar\int d^3\bfr{r} \,\,\lad^*(\bfr{r},t) \,\, \bfr{r} \times \nabla \lad(\bfr{r},t)}{\int dx \,\,|\lad(x,t)|^2},
 \end{aligned}
\end{equation}
which can then be written as 
\begin{equation}\label{eq:angl_mom_final}
 \begin{aligned}
 \tilde{\textbf{L}} = \frac{\braket{\phi|\hat{L}|\phi}}{\braket{\phi|\phi}},
 \end{aligned}
\end{equation}
with $\hat{L} = -i\hbar \bfr{r} \times \nabla$, 
which is the definition of the angular momentum in non relativistic quantum mechanics. 

\subsection{The external forces}\label{sec:forces}
The numerator of the external forces as defined in Eq. \eqref{eq:def_force} can be expressed in terms of the $(\lad,\lad^*)$ variables as follows 
\begin{equation}\label{eq:def_force_phi}
 \begin{aligned}
 \int dx &\psi(x,t)^2 \derivb{x}{}v(x) = \int dx |\phi(x)|^2 \derivb{x}{}v(x) \\
 &+  \frac{1}{2}\big(e^{-2i\omega_0t} (\phi(x))^2 + e^{+2i\omega_0t} (\phi^*(x))^2  \big) \derivb{x}{}v(x),
 \end{aligned}
\end{equation}
and as before, by performing the non relativistic limit together with a time average, the oscillating terms vanish and one obtains 
\begin{equation}\label{eq:def_force_phi_2}
 \begin{aligned}
 \lim_{c\rightarrow \infty}\int dx &\psi(x,t)^2 \derivb{x}{}v(x) = \int dx |\phi(x)|^2 \derivb{x}{}v(x). 
 \end{aligned}
\end{equation}
Inserting Eq. \eqref{eq:def_force_phi_2} in the definition of the forces (\textit{i.e.} Eq. \eqref{eq:def_force}) we obtain therefore,  
in the non relativistic limit, the usual expression from the Ehrenfest theorem 
\begin{equation}\label{eq:def_force_3}
 \tilde{F}[\psi,\dot{\psi}] = \frac{\braket{\phi|\hat{F}|\phi}}{\braket{\phi|\phi}},
\end{equation}
where $\hat{F}=-\derivb{x}{}v(x)$. 

\section{Non hermitian quantum mechanics and open systems}\label{sec:nh_qm}
In the present section we show how, using our classical framework, we can naturally recover the main features of 
the non hermitian description of open systems in quantum mechanics.  
Non-hermitian quantum systems is a subject \textit{per se} which goes beyond these simple derivations 
as it ranges from the treatment of real-valued spectrum Hamiltonian through imposing 
$\mathcal{PT}$ symmetry\cite{BenBroJon-AJP-03,BenHoo-RMP-24}, to the effective treatment of open systems 
in quantum mechanics\cite{BrePet-nhqm-02,RivHue-nhqm-12,RocPalCicBag-OSID-22}.  
We shall here work in the second case which focusses on the effective description of a quantum system 
coupled to a bath, which therefore gives rise to a dynamic governed by a non hermitian Hamiltonian. 

\subsection{Non hermitian quantum mechanics resulting from a classical friction force}
Coming back to the discrete classical phonon model, we simply assume now that the springs  
are submitted to a non-conservative force of the type $F=\alpha v$, which therefore typically results 
from the coupling of the system to an external bath with which it can exchange energy. 
As a consequence, we will be dealing with an open system, whose total energy is not conserved. 
If $\alpha<0$, it corresponds to a dissipative force, and on the contrary if $\alpha >0$ the system should gain energy. 
In the context of the phonon model, it naturally translates into an additional non-conservative force, labelled here $F_n^{\text{nc}}(t)$, for each spring indexed by $n$.
The expression of these forces is therefore simply $F_n^{\text{nc}}(t) = \alpha_n \dot{u}_n(t) $, 
where we recall that $\dot{u}_n(t)$ is the transverse velocity of the $n$-th mass, 
and where $\alpha_n\in \mathbb{R}$ is the friction coefficient which, in all generality, 
might depend on the index $n$ of the chain. 

In the continuous limit, the dissipative force becomes proportional to the continuous velocity field, and expressed 
in terms of the scaled variables $\dot{\psi}(x,t)$ it yields 
\begin{equation}\label{eq:friction}
 F^{\text{nc}}(x,t) = \alpha(x)\lambda_c \dot{\psi}(x,t)\sqrt{dx},
\end{equation}
where $\alpha(x)\in \mathbb{R}$ is the $x$ dependency of the friction coefficient, and $\sqrt{dx} = \lim_{a\rightarrow 0} \sqrt{a}$. 
As the Klein-Gordon is nothing but the Newtonian law of motion for the continuous phonon model, including the friction force simply 
yields 
\begin{equation}\label{eq:kg_nh}
 \derivb{t}{2} \psi(x,t) + \omega_0^2\big(1+2\frac{v(x)}{mc^2}\big)\psi(x,t) - c^{2}\derivb{x}{2}\psi(x,t) 
 = \frac{\alpha(x)}{m} \dot{\psi}(x,t),
\end{equation}
where now the right-hand side contains the non-conservative force (notice that to get to Eq. \eqref{eq:kg_nh} we divided 
the whole continuous Newtonian equation by $\lambda_c \sqrt{dx}$). 
As in Sec. \ref{sec:schro_smooth}, we can now switch to the $(\lad,\lad^*)$ variables to express the new Klein-Gordon equation of Eq. \eqref{eq:kg_nh}, 
which yields 
\begin{equation}\label{eq:kg_nh_2}
 \begin{aligned}
 &-i\hbar \derivb{t}{}{\lad}(x,t) -\frac{\hbar^2}{2m}\derivb{x}{2}\lad(x,t) 
+ \bigg(v(x)+ i\frac{1}{2}\frac{\hbar \alpha(x)}{m}\bigg)\lad(x,t) \\ & =\!\! e^{2i\omega_0 t}\!
  \bigg(\frac{\hbar^2}{2m}\derivb{x}{2}\lad^*(x,t) -(v(x) + i\frac{1}{2}\frac{\hbar \alpha(x)}{m}\lad^*(x,t)\bigg) .
 \end{aligned}
\end{equation}
We notice from Eq. \eqref{eq:kg_nh_2} that the additional term arising from the friction force (\textit{i.e.} those involving $\alpha(x)$) 
are multiplied by the purely imaginary number $i$. 
The latter originates from the fact that the associated friction force of Eq. \eqref{eq:friction} is proportional to $\dot{\psi}$ 
(we found a similar case in the case of the momentum in Sec. \ref{sec:momentum}). 
As done in Sec. \ref{sec:schro_smooth}, we now perform the non relativistic limit together with a short-time average 
such that all terms proportional to $e^{+2i\omega_0 t}$ vanish. Therefore, in the non relativistic limit, 
the right-hand side of Eq. \eqref{eq:kg_nh_2} vanishes, and we get the associated $\schro$ equation, 
which can be written as 
\begin{equation}\label{eq:sch_nh}
 i\hbar \derivb{t}{}{\lad}(x,t) = -\frac{\hbar^2}{2m}\derivb{x}{2}\lad(x,t) 
+ \big(v(x)+ iv^{\text{nc}}(x)\big)\lad(x,t),
\end{equation}
where we introduced the non-conservative potential as follows
\begin{equation}
 v^{\text{nc}}(x) = \frac{1}{2}\frac{\hbar \alpha(x)}{m}.
\end{equation}
Therefore, Eq. \eqref{eq:sch_nh} contains now a complex-valued potential which leads to a non-hermitian Hamiltonian. 
The imaginary potential in Eq. \eqref{eq:sch_nh} is therefore straightforwardly connected to a 
classical friction force through the complex valued 
change of variable $(\psi,\dot{\psi})\rightarrow (\lad,\lad^*)$. 

\subsection{The decay of the norm of the $\schro$ wave function as the power of dissipative forces}
As the system is submitted to dissipative forces, 
an interesting property to look at is the time variation of the total energy.  
The latter can be obtained straightforwardly by 
invoking a strictly classical argument: 
the time variation of the total mechanical energy of a classical system is simply given by the power of the non conservative forces acting on it. 
In the case of the discrete classical model, the power of the non conservative force is simply 
\begin{equation}
 \begin{aligned}
 \mathcal{P}[\mathbf{U}(t)] & = \sum_n F_n^{\text{nc}}(t) \dot{u}_n(t) \\
                             & = \sum_n \alpha_n \big(\dot{u}_n(t)\big)^2,\\
 \end{aligned}
\end{equation}
or expressed in terms of the scaled variables ${\mathbf{\Psi}(t)=\{ \psi_n(t)\}}$, as 
\begin{equation}
 \begin{aligned}
 \mathcal{P}[\mathbf{\Psi}(t)] & = \sum_n \alpha_n \big(\dot{\psi}_n(t)\lambda_c \sqrt{a}\big)^2,
 \end{aligned}
\end{equation}
which translates into the continuum limit into 
\begin{equation}
 \begin{aligned}
 \mathcal{P}[\psi] & = \lambda_c^2\int dx \alpha(x) \dot{\psi}(x,t)^2,
 \end{aligned}
\end{equation}
and therefore we obtain that 
\begin{equation}\label{eq:d_dt_h_nh}
 \frac{d}{dt} H[\psi,\dot{\psi}] = \lambda_c^2\int dx \alpha(x) \dot{\psi}(x,t)^2.
\end{equation}
Translated now into the $(\lad,\lad^*)$ variables and performing the non relativistic limit together with a short-time 
average, we obtain 
\begin{equation}
 \begin{aligned}\label{eq:power_phi}
 \mathcal{P}[\lad] & = \omega_0 \int dx 2 v^{\text{nc}}(x) |\lad(x,t)|^2.
 \end{aligned}
\end{equation}
Therefore we see that in the case of a dissipative system, corresponding to $v^{\text{nc}}(x)\le 0$, the total energy will necessarily 
decrease, while in the case of $v^{\text{nc}}(x) >0$, the total energy will increase. 
Now, as norm of the Klein-Gordon fields is simply the total energy $H[\psi]$ divided by $mc^2$, 
we can use Eq. \eqref{eq:norm_nr} and write the time derivative in the non relativistic limit as 
\begin{equation}
 \frac{d}{dt}\lim_{c\rightarrow \infty} \frac{H[\psi,\dot{\psi}]}{mc^2} = \frac{d}{dt}\mathcal{N}(t), 
\end{equation}
which, using Eq. \eqref{eq:d_dt_h_nh} and Eq. \eqref{eq:power_phi}, leads to 
\begin{equation}
 \frac{d}{dt}\mathcal{N}(t) = \frac{1}{\hbar} \int dx 2v^{\text{nc}}(x) |\lad(x,t)|^2, 
\end{equation}
and as $\mathcal{N}(t)$ is the $L^2$ norm of the $\schro$ wave function, we obtain 
\begin{equation}
 \frac{d}{dt}\langle \lad(t)|\lad(t)\rangle = \frac{1}{\hbar} \int dx 2v^{\text{nc}}(x) |\lad(x,t)|^2. 
\end{equation}
We thus recover the standard result of non relativistic quantum mechanics: the time derivative of the wavefunction norm 
is proportional to the expectation value of the imaginary part of the potential at a given time $t$. 
Also, as the norm of the wave function is tightly linked to the rest mass energy expressed in units of $mc^2$, 
the non conservation of the norm implies a non conservation of the mass of the system. 

\section{Conclusion}\label{sec:conclusion}
In the present work we have established a formal mapping between a phonon model 
made of classical particles and springs, and both the Klein-Gordon and $\schro$ equations. 
In that framework, the Klein-Gordon equation appears as the continuous limit of the Newtonian equation 
of motion for the transverse motion of this classical phonon model. 
On the other hand, the $\schro$ equation arises as the Newtonian equation of motion associated 
to the low-frequency envelope of these transverse oscillations, 
which is retrieved by a simple change of variable and a non relativistic limit in the Klein-Gordon equation. 
The complex numbers of the $\schro$ equation and wave function appear here as a specific choice 
of a complex-valued change of variables mixing positions and velocity, which is particularly well suited 
to describe the low-frequency envelope of the oscillations of this classical model. 
The introduction of a time-invariant scalar product on the space of real-valued Klein-Gordon field allows, 
with the previously mentioned complex-valued change of variables, 
to recover the usual $L^2$ scalar product of non relativistic quantum mechanics.  

Building on these results on the $\schro$ equation and wave function, the usual 
quantum expectation values (\textit{i.e.} Hamiltonian, position, linear and angular momentums) 
are then obtained using the global classical properties expressed in the complex-valued low-frequency representation. 
Last but not least, the introduction of a simple non conservative friction force in our classical model 
can be rewritten as a non unitary $\schro$ equation, as in non hermitian non relativistic quantum mechanics. 

Among the new features brought by this mapping between classical and quantum mechanics, 
we show here that the eigenvectors of the non relativistic Hamiltonian can then be understood 
as the continuous limit of the normal modes of the phonon model. 
Also, the norm of the $\schro$ wave function is directly proportional 
to the high-frequency energy density of this classical model, which is tightly linked to the rest mass energy. 
Another feature concerns the Laplacian term appearing in both the Klein-Gordon and $\schro$ equations,
which have a rather different classical interpretation with respect to the usual kinetic energy interpretation. 
Indeed, within our framework, the Laplacian describes a potential energy term representing 
the intrinsic incompressibility of the phonon system. 
The latter sheds a new light on the confinement energy or Heisenberg inequality 
in quantum mechanics stating that localization increases the kinetic energy: 
the increase in energy results here as the unavoidable increase in potential energy of the system induced 
by a tension in the phonon field.

Having established in the first part of this series how the $\schro$ equation can be expressed 
as the Newtonian formalism for this continuous classical phonon model, the second part\cite{Giner-heisenberg} of this series 
will focus on the classical Hamiltonian formalism and show how it allows to obtain the algebra 
of non commuting operators and the Heisenberg framework. 
This will also provide a deeper link with the Hilbert-space structure of Klein-Gordon fields brought by the 
Lorentz invariant inner product of Mostafazadeh\cite{Mostafazadeh-CQG-02,Mostafazadeh-AP-04,MosZam-AP-06},  
together with the various alternative formalisms proposed to reconcile the Klein-Gordon equation 
with the usual interpretation of quantum mechanics\cite{NewWig-RMP-49,FesVil-RMP-58,Trubenbacher-ZFN-89,BriEngSus-ZFN-91,Lammerzahl-JMP-93,Namsrai-IJTP-98}. 

Regarding the perspective of this work, 
looking at the Dirac equation seems to be a natural route as the 
latter is tightly linked to the Klein-Gordon equation. Also, the present formalism does not allow to introduce the notion of particles, 
which intrinsically results from a quantization of the Klein-Gordon (or Dirac) field. 
We aim to investigate this latter path in future works.

\appendix
\section{Main results associated with the real-valued Klein-Gordon fields}
\label{seq:schro_from_kg_usual}
This annex provides a brief summary of the main results associated to real-valued Klein-Gordon fields which are 
useful for this work. 
We begin in Sec. \ref{sec:kg_cst_annex} by recalling how the Klein-Gordon equation is usually retrieved and 
the main quantities, such as the energy or momentum, which are related to these fields. 
Then in Sec. \ref{sec:kg_usual} we recall how the $\schro$ is obtained from the Klein-Gordon equation, 
and briefly discuss the reason of why we do not follow this path. 
\subsection{The Klein-Gordon equation and its constants of motion}
\label{sec:kg_cst_annex}
The Klein-Gordon equation is usually derived starting with the relativistic energy relation 
$E^2=p^2c^2+m^2c^4$, which is then quantized using the quantum equivalence relations between observables and operators, 
\textit{i.e.} $(E,p) \rightarrow(\hat{E}\equiv i\hbar \derivb{t}{},-i\hbar \nabla)$. 
As one deals with spatial and time differential operators, one needs to introduce a scalar field $\psi(\bfr{r},t)$ 
on which such operators act, and this procedure leads then to the following equation of motion for $\psi(\bfr{r},t)$ 
\begin{equation}\label{eq:KG_1}
 \derivb{t}{2}\psi(\bfr{r},t) -c^2 \Delta \psi(\bfr{r},t) + \omega_0^2 \psi(\bfr{r},t) = 0, 
\end{equation}
where $\omega_0$ is defined in Eq. \eqref{eq:omega_0}. 
Although this straightforward derivation can be satisfactory from the point of view of connecting special relativity 
and quantum mechanics, it nevertheless relies on the introduction of a field $\psi(\bfr{r},t)$, 
potentially complex-valued and whose interpretation is not straightforward 
($|\psi(\bfr{r},t)|^2$ cannot be interpreted as the usual probability density as its $L^2$-norm is not constant in time), 
together with differential operators borrowed from non relativistic quantum mechanics. 

In the context of classical field theory, the real-valued solutions of the Klein-Gordon equation are usually obtained 
as the stationary conditions of the action corresponding to the following Lagrangian density 
\begin{equation}\label{eq:lagrange_qft}
 \mathcal{L}(\psi(\bfr{r},t)) = \big(\derivb{t}{}\psi(\bfr{r},t)\big)^2 - c^2\big(\nabla\psi(\bfr{r},t)\big)^2 - \omega_0^2 \big(\psi(\bfr{r},t)\big)^2.
\end{equation}
The form of this Lagrangian is designed to be Lorentz invariant, \textit{i.e.} transforming 
as a scalar through the symmetry operations of special relativity. 
One can then apply Noether's theorem to the Lorentz 
group and obtain the constants of motion associated to the symmetry operations of this group, \textit{i.e.} 
boosts and rotations. 
The total energy defined by 
\begin{equation}\label{eq:cl_htot_qft}
 H[\psi,\dot{\psi}] = \int d\bfr{r}\varepsilon(\psi,\dot{\psi},\bfr{r}),
\end{equation}
where the energy density is given by 
\begin{equation}
 \varepsilon(\psi,\dot{\psi},\bfr{r}) = \frac{1}{2}\big(\derivb{t}{}\psi(\bfr{r},t)\big)^2 + \frac{1}{2}\omega_0^2 \big(\psi(\bfr{r},t)\big)^2 
                         + \frac{c^2}{2}  \big|\nabla\psi(\bfr{r},t)\big|^2,
\end{equation}
is a result of the time component of the conserved vector (\textit{i.e.} the $T^{00}$ element of the energy-momentum tensor) associated to the continuous boost symmetry. 
Also, the total momentum defined as follows 
\begin{equation}\label{eq:mom}
 \bfr{P} = -\int d\bfr{r}^3 \derivb{t}{}\psi(\bfr{r},t) \nabla \psi(\bfr{r},t),
\end{equation}
appears as a result of the spatial component of the same energy-momentum tensor (\textit{i.e.} $T^{0i}$), 
while the angular momentum given by 
\begin{equation}\label{eq:ang_mom}
 \bfr{L} = \int d\bfr{r}^3 \derivb{t}{} \psi(\bfr{r},t) \,\,\bfr{r} \times \nabla \psi(\bfr{r},t),
\end{equation}
is a result of rotational invariance of the system. 
One can also show the following Ehrenfest relations\cite{ReiFer-PRL-91}  
\begin{equation}\label{eq:ehren}
 \frac{d}{dt}\bfr{R}(t) = \frac{\bfr{P}}{m},
\end{equation}
where $\bfr{R}(t)$ is the associated energy barycentre
\begin{equation}
 \bfr{R}(t) = \int d\bfr{r}^3 \,\,\bfr{r} \,\,\varepsilon(\psi,\dot{\psi},\bfr{r}).
\end{equation}
Eq. \eqref{eq:ehren} can be obtained by straightforward algebra\cite{ReiFer-PRL-91} or by symmetry considerations (see 
for instance Ref. \onlinecite{QFTong}). 

These results on real-valued Klein-Gordon fields are obtained in Sec. \ref{sec:frenkont} based on the classical 
nature of the phonon model. 

\subsection{The usual path connecting the Klein-Gordon and $\schro$ equation}
\label{sec:kg_usual}
Coming now to the usual way one one obtains the $\schro$ equation from the Klein-Gordon equation 
(see for instance Ref. \onlinecite{Greiner-book}), it is done by writing the Klein-Gordon field as 
\begin{equation}\label{eq:psi_kg_usual}
 \psi(\bfr{r},t) = e^{-i\omega_0 t}f(\bfr{r},t), 
\end{equation}
where $f(\bfr{r},t)$ typically corresponds to the envelope of the oscillations of $\psi(\bfr{r},t)$, 
and in the limit where the time variations of $f(\bfr{r},t)$ are small on the scale of $\omega_0^{-1}$, \textit{i.e.} 
\begin{equation}\label{eq:slow_1}
 \bigg|\frac{\derivb{t}{}f(x,t)}{f(x,t)}\bigg| \ll \omega_0,
\end{equation}
the complex-valued field $f(\bfr{r},t)$ satisfies the $\schro$ equation,
\begin{equation}
 i\hbar \derivb{t}{}f(\bfr{r},t) = -\frac{\hbar^2}{2m} \Delta f(\bfr{r},t). 
\end{equation}
Nevertheless, assuming the form of Eq. \eqref{eq:psi_kg_usual} together with fulfilling the Klein-Gordon equations 
implies that the Klein-Gordon field is intrinsically complex, which exclude the real-valued solutions. 
The latter aspect is crucial in the context of our derivations as the present work relies on a Newtonian interpretation 
of the Klein-Gordon equation, which therefore relies on a real-valued field $\psi(\bfr{r},t)$. 
Therefore, although the complex-valued form of Eq. \eqref{eq:psi_kg_usual} is perfectly legitimate 
from a mathematical perspective, 
it is not a desirable path for our derivations which aim at connecting the mathematical formulation of 
quantum mechanics with classical mechanics.

\bibliography{biblio}
 \end{document}